\makeatletter\def\input@path{{./}}\makeatother
\documentclass[pdflatex,sn-nature]{sn-jnl}%

\usepackage{graphicx}%
\usepackage{multirow}%
\usepackage{amsmath,amssymb,amsfonts}%
\usepackage{amsthm}%
\usepackage{mathrsfs}%
\usepackage[title]{appendix}%
\usepackage{xcolor}%
\usepackage{textcomp}%
\usepackage{manyfoot}%
\usepackage{booktabs}%
\usepackage{array}%
\usepackage[skip=2pt,font=small]{subcaption}%
\DeclareCaptionFont{panelletter}{\sffamily\bfseries\fontsize{8}{10}\selectfont}
\usepackage{algorithm}%
\usepackage{algorithmicx}%
\usepackage{algpseudocode}%
\usepackage{listings}%
\usepackage{tikz}%
\usetikzlibrary{positioning, arrows.meta, calc, fit, backgrounds, decorations.pathreplacing}%

\definecolor{QdBackbone}{HTML}{1d4ed8}%
\definecolor{QdQuantum}{HTML}{ea580c}%
\definecolor{QdAdapter}{HTML}{16a34a}%
\definecolor{QdSoft}{HTML}{eff6ff}%

\theoremstyle{thmstyleone}%
\newtheorem{theorem}{Theorem}%

\theoremstyle{thmstyletwo}%

\theoremstyle{thmstylethree}%

\graphicspath{{./}}

\begin{document}

\author[1,2]{\fnm{Xiaoqiang} \sur{Wang}}

\author[3]{\fnm{Mengyang} \sur{Xiong}}

\author*[1,2]{\fnm{Jun} \sur{Dai}}\email{jun.dai@mila.quebec}

\author*[1,2]{\fnm{Bang} \sur{Liu}}\email{bang.liu@umontreal.ca}

\affil[1]{\orgdiv{D\'epartement d'informatique et de recherche op\'erationnelle \& Institut Courtois}, \orgname{Universit\'e de Montr\'eal}, \orgaddress{\city{Montr\'eal}, \state{QC}, \country{Canada}}}

\affil[2]{\orgname{Mila, Quebec AI Institute}, \orgaddress{\city{Montr\'eal}, \state{QC}, \country{Canada}}}

\affil[3]{\orgname{McGill University}, \orgaddress{\city{Montr\'eal}, \state{QC}, \country{Canada}}}

\title{Circuit hypernetworks for quantum-augmented diffusion language models}

\abstract{
Improving pretrained language models with limited adaptation data calls for computational modules that are both effective and scalable. Parameterized quantum circuits offer expressive transformations for language modeling, yet existing quantum-enhanced language models apply the same predefined circuit design to every token in an input sequence, regardless of its content. This limitation is particularly acute in masked-diffusion language models, where token representations evolve across iterative denoising steps while the circuit design remains fixed.
We introduce HyperQ, a quantum-augmented masked-diffusion language model that dynamically synthesizes token-specific quantum circuits. Built upon an IQP-inspired architecture, HyperQ employs a lightweight circuit hypernetwork to generate rotation angles and measurement bases conditioned on each token's hidden state. The resulting quantum expectation values are projected back into the frozen language-model backbone through a residual connection. To support wider quantum registers, HyperQ bounds circuit interaction degree, maintaining per-qubit gradient scale and readout cost independent of register width.
Across six zero-shot benchmarks, scaling HyperQ from 16 to 64 qubits improves average accuracy from 47.65 to 54.30, surpassing the frozen backbone's 49.59 by 4.71 points while using only one tenth of the fine-tuning data used by classical baselines. We further execute the generated circuits on an IBM 156-qubit superconducting quantum processor, demonstrating their feasibility on physical quantum hardware.
}

\keywords{quantum machine learning, diffusion language models, circuit
hypernetworks, variational quantum circuits,
hybrid quantum--classical computing}

\maketitle

\section{Introduction}
\label{sec:intro}

Large language models (LLMs) predominantly use the attention-based transformer architecture~\cite{grattafiori2024llama3, yang2024qwen2, guo2025deepseek, vaswani2017attention}, and their progress has relied primarily on larger models, training corpora and amounts of classical compute~\cite{kaplan2020scaling, hoffmann2022training, snell2024scaling}. These investments support both autoregressive models, which factorise text from left to right~\cite{radford2019language}, and diffusion-based models, which iteratively refine corrupted representations~\cite{ho2020ddpm}. Masked-diffusion models extend diffusion to text by denoising masked positions in parallel under bidirectional context~\cite{austin2021d3pm, li2022diffusionlm, yu2025ddsurvey}. This family now includes billion-parameter open models~\cite{nie2025llada, ye2025dream}, models pre-trained on 12T tokens~\cite{nie2026illada} and 100B-scale mixtures of experts~\cite{bie2025llada2, zhu2026lladamoev2}. Diffusion models can also be competitive when data, rather than compute, is the limiting resource~\cite{prabhudesai2025diffusionbeatsar}. %
Masked diffusion repeatedly updates token representations as the surrounding context changes. This provides a natural setting for learning transformations whose parameters depend on the current representation. We investigate whether quantum circuits can provide useful transformations of this kind within a frozen diffusion language model.

Parameterised quantum circuits offer one source of such capacity because their trainable rotation angles can define learnable computations inside a transformer~\cite{biamonte2017qml, cerezo2021vqa}. Existing integrations modify language models through several related mechanisms. Variational circuits have been inserted into a 150M-parameter transformer, where they match classical quality~\cite{kong2025hyqut}, and into self-attention, where a classical bottleneck of the same size matches the quantum layer~\cite{chen2025qasa}. Another line of work constrains model updates through unitary structure.
Cayley unitary adapters improve a frozen 8B model by about $1.4\%$ in perplexity~\cite{singh2026cayley}, and related unitary updates pursue parameter efficiency~\cite{koikeakino2025qpeft}. Hybrid quantum-tensor and quantum low-rank adapters provide a further route to adapting a frozen language model~\cite{kong2025qtha, xing2026qlora, raj2025quic}.

These integrations fix one circuit design before training and apply it to every token, regardless of its content. In masked diffusion, the same fixed circuit is therefore applied to a succession of different representations at each position.
Quantum architecture search adapts neural architecture search to choose gates and connectivity, including through differentiable relaxations of discrete choices~\cite{wang2022quantumnas, zoph2017nas, zhang2022dqas, kumar2025rhodarts}. Continuous synthesis and analytical design criteria instead optimise or rank circuits for a fixed task or target~\cite{singh2026synthesis, jing2026certified, lan2026dla, roseler2026expressible}. FlowQNet samples discrete gate sequences with a trajectory-balance objective that preserves probability mass across comparable terminal architectures, but each sampled architecture still requires fresh optimisation of its continuous parameters~\cite{dai2025flowqnet}. These methods generally operate at relatively small qubit counts and independently of the language model, so they do not emit and train a distinct circuit directly from each evolving token representation. 
Unlike these searches, which choose one circuit for a whole task, HyperQ generates a circuit for each token from that token's representation. All of these circuits share one gate template, with a single-qubit rotation on every qubit and a two-qubit rotation on every edge of a bounded-degree graph. A hypernetwork, trained jointly with the language model, maps each token representation to every rotation angle of this template and to a measurement basis for every qubit. A rotation whose angle is close to zero acts nearly as the identity, so these angles determine which gates effectively act on the token and how strongly, and we call the result the token's effective circuit. To separate this continuous generation from discrete selection, we also compare HyperQ with a token-conditioned architecture search that selects whole gate groups for each token.

HyperQ implements the hypernetwork, a network that emits the parameters of another module~\cite{ha2016hypernetworks}, as a low-rank adapter~\cite{hu2021lora}. Each token's circuit returns a quantum readout, which is projected into the attention computation. We train the hypernetwork and output projections while keeping the pretrained model frozen.

Emitting parameters for every token requires a structured circuit family whose parameters can vary with the hidden state while its evaluation remains tractable. HyperQ restricts its emitted circuits to the Instantaneous Quantum Polynomial (IQP) family~\cite{bremner2011iqp, bremner2017iqp}. Each circuit applies Hadamard gates before and after a commuting diagonal block. Within this block, the rotation angles $\alpha$ parameterise single-qubit Z-rotation gates $R_z$, while the two-qubit rotation angles $\omega$ parameterise two-body ZZ-coupling gates $R_{zz}$ on a fixed edge set $E$. The edge set forms a ring layer and a chord layer, giving every qubit degree four at 16, 32 and 64 qubits. After the IQP block, $\psi$ parameterises each qubit's measurement axis. For qubit $q$, HyperQ returns the expectation value $\langle Z_q\rangle$ of the Pauli-Z observable $Z_q$ to the transformer block.

The fixed degree interaction graph makes these token-specific circuits affordable as their width increases. For a circuit width of $n$ qubits, unstructured variational circuits can exhibit barren plateaus, with gradient variance exponentially small in $n$~\cite{larocca2025barren}. Exact simulation of a generic $n$-qubit pure state also requires a statevector containing $2^n$ amplitudes, making repeated per-token evaluation prohibitively expensive inside a billion-parameter model. Circuit structure links these two difficulties because trainable circuit families can remain efficiently classically simulable~\cite{cerezo2025simulability}. In HyperQ, the two-body generators yield a closed-form expression for each expectation value, so all $n$ readouts can be evaluated exactly with total cost $\Theta(n)$ and without constructing the full statevector. 
Under the specified independent reference ensemble, the variances of derivatives with respect to the local rotation and incident coupling parameters depend on interaction degree. They remain independent of circuit width when this degree is fixed.
Mixtures of IQP circuits can be classically efficiently trained~\cite{lee2026mixediqp}. %
HyperQ's result applies specifically to its circuit family, reference initialisation and per-token expectation readout.

At $64$ qubits, HyperQ raises the six-benchmark average from $49.59$ for the frozen $1.1$ billion parameter model to $54.30$. It also exceeds the classical low-rank-adapted model by $3.67$ points. HyperQ achieves these results with $20{,}000$ fine tuning pairs, compared with $200{,}000$ for the fine tuned classical baselines. Performance improves across the three evaluated circuit widths. At $16$, $32$ and $64$ qubits, generating circuit parameters from token representations outperforms the tested fixed circuits and token-conditioned discrete motif searches. The hardware evaluation also retains most of the performance obtained with analytic readouts. On the paired $64$-qubit evaluation subset, superconducting-device measurements reduce the average score by $1.12$ points with the model weights held fixed. These findings establish context-dependent circuit generation as an effective approach to adapting a frozen diffusion language model. The combination of learned circuit parameters and exactly computable readouts makes it possible to study circuit based computation within a billion parameter model. Future extensions could use shared circuits to coordinate token updates during denoising and conditional IQP circuits to sample the final output through the Born rule. This would connect language generation to a broader research goal, combining efficient classical training with useful joint output distributions that are hard to sample classically.

\begin{figure}[!tp]
\centering
\includegraphics[width=\textwidth]{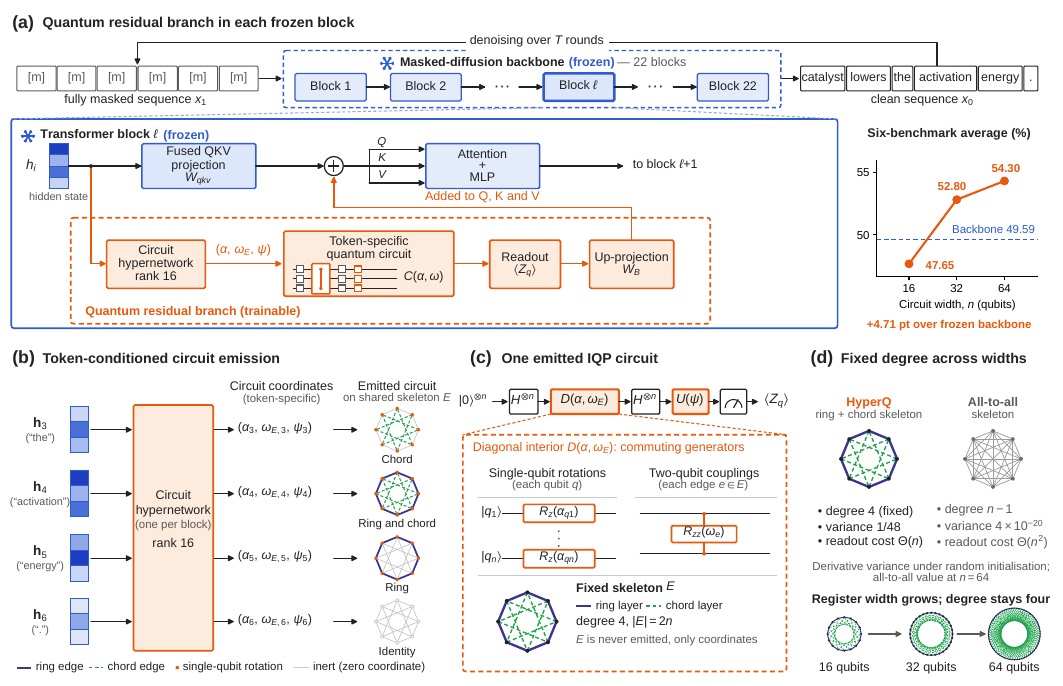}
    \caption{\textbf{Token-conditioned quantum circuit emission in HyperQ.}
    \textbf{(a)}, In each of the 22 frozen blocks of the masked-diffusion
    backbone~\cite{nie2025llada}, a trainable quantum branch reads a token's
    hidden state, emits and executes that token's circuit, and adds the
    projected readout to $Q$, $K$ and $V$. Right, six-benchmark average
    against register width. Dashed line, the frozen backbone at 49.59.
    \textbf{(b)}, One hypernetwork per block maps each token to its own
    coordinates $(\alpha,\omega_E,\psi)$ on a shared skeleton $E$, where a zero
    coordinate removes its gate.
    \textbf{(c)}, One emitted IQP circuit~\cite{bremner2011iqp}, represented by
    its interaction graph. Qubits are arranged on a ring, with vertices
    encoding single-qubit rotations $R_z(\alpha_q)$ and ring-plus-chord edges
    encoding two-qubit couplings $R_{zz}(\omega_e)$. This representation
    specifies the diagonal block while directly exposing the fixed-degree
    connectivity that enables tractable widening.
    \textbf{(d)}, Because $\langle Z_q\rangle$ factorises over the edges at $q$,
    fixed degree four keeps the readout cost at $\Theta(n)$ and, under reference
    initialisation, the derivative variance at $1/48$ ($4\times10^{-20}$ for
    an all-to-all skeleton at 64 qubits).}
    \label{fig:overview}
\end{figure}

\section{Results}\label{sec:results}

In this section, we first describe HyperQ and its integration into a frozen diffusion language model. We then evaluate its performance on six language benchmarks against classical models and other quantum based adaptations. Comparisons with fixed circuits, discrete circuit searches and circuit ensembles assess the benefit of predicting continuous circuit parameters for each token. Next, we examine how the learned couplings and single-qubit measurement bases vary across tokens and lexical categories. We study accuracy and evaluation cost at $16$, $32$ and $64$ qubits, alongside the analytical properties of the local readouts. Finally, we evaluate performance with an IBM quantum computer and examine how HyperQ preserves parallel decoding across different generation orders.

\subsection{Token-conditioned circuits in a frozen diffusion model}\label{subsec:framework}

HyperQ generates circuit parameters from each token’s current representation (Fig.~\ref{fig:pipeline}). In each transformer block, a small hypernetwork reads the input to the query, key and value projection. %
A hypernetwork is a neural network whose outputs are the parameters of another model~\cite{ha2016hypernetworks}. In HyperQ it consists of a layer normalisation, a projection to $r=16$ dimensions, one non-linearity and three linear output heads (Fig.~\ref{fig:hyper}). For each token, the heads output the parameters of that token's quantum circuit, namely one single-qubit rotation angle per qubit, one two-qubit rotation angle per two-qubit gate and one measurement basis per qubit. The circuit returns local expectation values $\langle Z_q\rangle$. These values are projected and added to the query, key and value tensors $Q$, $K$ and $V$. The diffusion model remains frozen while the hypernetwork and readout projections are trained.

Each of the $22$ transformer blocks contains a quantum residual branch that receives the hidden state $h$ of each token and generates the parameters of that token's circuit. All tokens share one gate template, which sets the gates that can act. For an $n$-qubit register, it consists of a Hadamard layer, an $R_z$ gate on every qubit, an $R_{zz}$ gate on every edge of an edge set $E$, a second Hadamard layer and a single-qubit basis rotation on every qubit before its $Z$ measurement. The edge set $E$ comprises a ring layer that joins each qubit to its two neighbours and a chord layer that joins qubit $q$ to qubit $q+c$ modulo $n$ for a fixed stride $c$. Every qubit therefore takes part in exactly four two-qubit gates, and $|E| = 2n$. This bounded degree gives every readout a closed form whose cost and gradient variance do not grow with the register width (Methods). The hypernetwork generates every angle of this template and every measurement basis for each token. Because a gate with angle zero is the identity, a generated angle close to zero makes its gate act nearly as the identity. The generated parameters therefore set which gates of the template effectively act on a token and how strongly. We call the result the token's effective circuit.

The generated parameters comprise single-qubit rotation angles $\alpha=\{\alpha_q\}$, two-qubit rotation angles $\omega_E=\{\omega_e:e\in E\}$ and measurement bases $\psi=\{\psi_q\}$, where $q$ indexes qubits and $e$ indexes edges. Each $\psi_q$ is a unit vector on the Bloch sphere with two free angles. It sets the single-qubit rotation applied to qubit $q$ just before its $Z$ measurement, so that qubit $q$ is measured along $\psi_q$. They specify a strict instantaneous quantum polynomial-time (IQP) block with a Hadamard layer, a diagonal interior of $R_z$ rotations and $R_{zz}$ couplings, a second Hadamard layer and a readout rotation $U(\psi)$. The denoising process, branch placement and emitted coordinates are shown in Fig.~\ref{fig:pipeline}a, Fig.~\ref{fig:pipeline}b and Fig.~\ref{fig:pipeline}c, respectively. The circuit returns the expectation value $\langle Z_q\rangle$, which a low-rank projection maps into the model dimension and adds to the query, key and value tensors $q,k,v$ produced by the fused projection $W_{qkv}$. Here, $q$ denotes the query tensor rather than the qubit index in $Z_q$.

\begin{figure}[!tp]
\noindent\makebox[\textwidth][l]{
\begin{tikzpicture}[
  >={Latex[length=1.5mm]},
  font=\fontsize{7}{8}\selectfont,
  tok/.style={draw=black!55, minimum width=0.30cm, minimum height=0.30cm,
              inner sep=0pt, font=\fontsize{5.2}{6}\selectfont},
  mtok/.style={tok, fill=black!72, text=white},
  ktok/.style={tok, fill=white},
  gtok/.style={tok, fill=QdAdapter!28, draw=QdAdapter},
  bx/.style={rounded corners=2pt, align=center, inner sep=1.6pt, thin,
             font=\fontsize{7}{8}\selectfont},
  cls/.style={bx, draw=QdBackbone, fill=QdSoft},
  qnt/.style={bx, draw=QdQuantum, fill=QdQuantum!18},
  obox/.style={bx, draw=QdAdapter, fill=QdAdapter!16},
  sm/.style={draw=black!60, fill=white, circle, inner sep=0pt, minimum size=2.9mm,
             font=\fontsize{5.6}{6}\selectfont},
  gate/.style={draw=QdQuantum, fill=QdQuantum!15, minimum size=0.26cm,
               inner sep=0pt, font=\fontsize{5}{5.5}\selectfont},
  fl/.style={->, semithick, QdBackbone!85},
  ql/.style={->, semithick, QdQuantum!85},
  tl/.style={->, semithick, black!55},
  rl/.style={->, thin, black!45},
  zoom/.style={draw=black!35, densely dashed, line width=0.35pt},
]

\begin{scope}[on background layer]
  \node[rounded corners=2pt, fill=QdBackbone!8, draw=QdBackbone!45, thin,
        fit={(1.14,2.56) (3.28,3.04)}, inner sep=0pt] {};
\end{scope}
\foreach \i in {0,...,4} { \node[mtok] at ({1.84+0.34*\i},4.80) {\texttt{[m]}}; }
\node[ktok] at (1.84,3.50) {$w_1$};
\foreach \i in {1,2,3,4} { \node[mtok] at ({1.84+0.34*\i},3.50) {\texttt{[m]}}; }
\foreach \i/\l in {0/{$w_1$},2/{$w_3$}} { \node[ktok] at ({1.84+0.34*\i},2.80) {\l}; }
\foreach \i in {1,3,4} { \node[mtok] at ({1.84+0.34*\i},2.80) {\texttt{[m]}}; }
\foreach \i/\l in {0/{$w_1$},2/{$w_3$},4/{$w_5$}} { \node[ktok] at ({1.84+0.34*\i},2.10) {\l}; }
\foreach \i in {1,3} { \node[mtok] at ({1.84+0.34*\i},2.10) {\texttt{[m]}}; }
\foreach \i/\l in {0/{$w_1$},1/{$w_2$},2/{$w_3$},3/{$w_4$},4/{$w_5$}}
  { \node[gtok] at ({1.84+0.34*\i},0.72) {\l}; }
\node[anchor=east] at (1.62,4.80) {$x_1$};
\node[anchor=east] at (1.62,3.50) {$x_{t+1}$};
\node[anchor=east, QdBackbone] at (1.62,2.80) {$x_{t}$};
\node[anchor=east] at (1.62,2.10) {$x_{t-1}$};
\node[anchor=east] at (1.62,0.72) {$x_0$};
\node[black!55] at (2.52,4.16) {$\vdots$};
\node[black!55] at (2.52,1.44) {$\vdots$};
\draw[tl] (0.34,4.92) -- (0.34,0.62);
\node[rotate=90, anchor=south, black!60] at (0.16,2.77) {denoise over $T$ rounds};

\node[QdBackbone] at (6.52,0.40) {$x_t$};
\draw[rounded corners=2.5pt, draw=QdBackbone, fill=QdBackbone!3]
  (4.86,0.72) rectangle (8.28,4.74);
\node[anchor=north west, QdBackbone] at (4.94,4.66) {$\times L$};
\node[cls, minimum width=1.24cm, minimum height=0.44cm] (wq) at (5.82,1.55)
  {$W_{qkv}$};
\node[qnt, minimum width=1.32cm, minimum height=0.44cm] (qb) at (7.32,1.55)
  {quantum\\residual};
\node[sm] (s0) at (6.52,2.25) {$+$};
\node[cls, minimum width=2.10cm, minimum height=0.34cm] (att) at (6.52,2.80)
  {self-attention};
\node[anchor=west, black!65] at (6.70,2.50) {$q,k,v$};
\node[sm] (s1) at (6.52,3.35) {$+$};
\node[cls, minimum width=2.10cm, minimum height=0.34cm] (mlp) at (6.52,3.90)
  {MLP};
\node[sm] (s2) at (6.52,4.45) {$+$};
\node[QdAdapter!75!black] at (6.52,5.02) {$p_\theta(\cdot\mid x_t)$};
\draw[fl] (6.52,0.55) -- (6.52,0.95);
\draw[QdBackbone!85, semithick] (5.82,0.95) -- (7.32,0.95);
\fill[black!45] (6.52,0.95) circle (0.9pt);
\draw[fl] (5.82,0.95) -- (5.82,1.33);
\draw[fl] (7.32,0.95) -- (7.32,1.33);
\draw[QdBackbone!85, semithick] (5.82,1.77) -- (5.82,2.25);
\draw[fl] (5.82,2.25) -- (6.37,2.25);
\draw[QdQuantum!85, semithick] (7.32,1.77) -- (7.32,2.25);
\draw[ql] (7.32,2.25) -- (6.67,2.25);
\draw[fl] (6.52,2.40) -- (6.52,2.63);
\draw[fl] (6.52,2.97) -- (6.52,3.20);
\draw[fl] (6.52,3.50) -- (6.52,3.73);
\draw[fl] (6.52,4.07) -- (6.52,4.30);
\draw[fl] (6.52,4.60) -- (6.52,4.88);
\draw[black!45, thin] (6.52,0.95) -- (5.06,0.95);
\draw[rl] (5.06,0.95) -- (5.06,3.35) -- (6.37,3.35);
\fill[black!45] (6.52,3.62) circle (0.9pt);
\draw[black!45, thin] (6.52,3.62) -- (8.04,3.62);
\draw[rl] (8.04,3.62) -- (8.04,4.45) -- (6.67,4.45);

\node[cls, minimum width=2.90cm, minimum height=0.30cm,
      font=\fontsize{6}{7}\selectfont] (h) at (10.97,0.42) {hidden state $h$};
\draw[rounded corners=2.5pt, draw=QdQuantum!70, fill=QdQuantum!4]
  (10.12,0.85) rectangle (11.82,3.70);
\draw[rounded corners=2.5pt, draw=QdQuantum!70, densely dashed, fill=QdQuantum!4]
  (10.12,3.78) rectangle (11.82,4.22);
\foreach \x in {10.37,10.77,11.17,11.57}
  { \draw[black!55, semithick] (\x,1.12) -- (\x,4.18); }
\node[black!60, font=\fontsize{5.6}{6}\selectfont, inner sep=1pt] at (10.97,0.98)
  {$|0\rangle^{\otimes n}$};
\foreach \x in {10.37,10.77,11.17,11.57} { \node[gate] at (\x,1.40) {$H$}; }
\foreach \x/\lb in {10.37/{$\alpha_1$},10.77/{$\alpha_2$},
                    11.17/{$\alpha_3$},11.57/{$\alpha_4$}}
  { \node[gate, minimum height=0.30cm] at (\x,1.95) {\lb}; }
\foreach \a/\b/\y in {10.37/10.77/2.35, 11.17/11.57/2.35,
                      10.77/11.17/2.62, 10.37/11.57/2.89} {
  \draw[QdQuantum, semithick] (\a,\y) -- (\b,\y);
  \fill[QdQuantum] (\a,\y) circle (1.0pt);
  \fill[QdQuantum] (\b,\y) circle (1.0pt);
}
\foreach \x in {10.37,10.77,11.17,11.57} { \node[gate] at (\x,3.42) {$H$}; }
\foreach \x/\lb in {10.37/{$u_1$},10.77/{$u_2$},11.17/{$u_3$},11.57/{$u_4$}}
  { \node[gate, minimum height=0.28cm] at (\x,4.00) {\lb}; }
\draw[QdQuantum!85, semithick] (12.10,0.57) -- (12.10,4.00);
\foreach \y in {1.95,2.62,4.00} { \draw[ql] (12.10,\y) -- (11.88,\y); }
\foreach \y/\lb in {1.95/{$\alpha$},2.62/{$\omega_{E}$},4.00/{$\psi$}}
  { \node[QdQuantum!85!black, font=\fontsize{6}{7}\selectfont, inner sep=1pt,
          anchor=west] at (12.16,\y) {\lb}; }
\draw[ql] (10.97,4.22) -- (10.97,4.47);
\node[obox, minimum width=2.90cm, minimum height=0.30cm,
      font=\fontsize{6}{7}\selectfont] (zq) at (10.97,4.62)
  {$\langle Z_q\rangle$ readout};

\draw[zoom] (3.28,3.04) -- (4.86,4.74);
\draw[zoom] (3.28,2.56) -- (4.86,0.72);
\draw[zoom] (7.98,1.77) -- (9.52,4.77);
\draw[zoom] (7.98,1.33) -- (9.52,0.27);

\node[font=\sffamily\bfseries\fontsize{8}{10}\selectfont, inner sep=0pt, anchor=west] at (-0.374,5.34) {(a)};
\node[font=\sffamily\bfseries\fontsize{8}{10}\selectfont, inner sep=0pt, anchor=west] at (4.86,5.34) {(b)};
\node[font=\sffamily\bfseries\fontsize{8}{10}\selectfont, inner sep=0pt, anchor=west] at (9.52,5.34) {(c)};
\end{tikzpicture}

}
\caption{The HyperQ pipeline at three scales, each panel read from bottom to
top. \textbf{(a)}, Masked-diffusion generation. A fully masked sequence $x_1$ is
denoised over $T$ rounds until the clean sequence $x_0$ remains, each round committing
the most confident predictions and re-masking the rest. \textbf{(b)}, One round,
expanded from the highlighted state $x_t$, returning the token distribution
$p_\theta(\cdot\mid x_t)$ of equation \eqref{eq:mdm}. Inside each block the fused query-key-value
projection is the only modified point, where a quantum branch runs in parallel
with the classical weight $W_{qkv}$ and is added into $q,k,v$ as a
residual. \textbf{(c)}, How that branch emits its circuit. From the hidden state $h$ of one
token the branch emits the rotation angles $\alpha$, the two-body couplings
$\omega_{E}$ carried on an edge set $E$ fixed once per register width, and the
per-qubit axes $\psi$ along which the strict IQP block of equation
\eqref{eq:iqp} is read out. The $\langle Z_q\rangle$ readout returns as the
residual that panel~(b) adds into $q,k,v$. Four of the $n$ wires are
drawn.}
\label{fig:pipeline}
\end{figure}

\subsection{Classical baseline comparison}\label{subsec:benchmarks}

We evaluated HyperQ on ARC-e, HellaSwag, PIQA, BoolQ, RACE and GSM8K, the union of the benchmarks reported by the compared methods (Table~\ref{tab:main}). HyperQ uses a tenth of the fine-tuning tokens used by the baselines. At 64 qubits, its six-benchmark average is $54.30$, compared with $53.79$ for Qwen3-8B, $53.37$ for LLaDA-8B and $49.59$ for the frozen LLaDA-1.1B backbone. HyperQ does not lead on WikiText perplexity, where lower is better, as Qwen3-8B reaches $10.61$ compared with $10.95$ for HyperQ at 64 qubits. HyperQ thus has the highest accuracy on the six downstream benchmarks among the evaluated classical models, including models up to eight times its size.

To separate the quantum residual branch from the effect of adding low-rank parameters, we compared HyperQ with a size-matched classical low-rank adapter. This adapter raises TinyLlama from $48.73$ to $49.98$ and LLaDA from $49.59$ to $50.63$, contributing roughly a point and a quarter in both cases. HyperQ at 64 qubits exceeds the adapted diffusion backbone by $3.67$ points. Because this control matches the classical component that HyperQ replaces, the remaining difference separates circuit emission from the gain obtained by adding parameters through a low-rank branch.

Earlier quantum-augmented language models do not provide a shared suite for direct downstream comparison. HyQuT~\cite{kong2025hyqut} reports generation quality on a dialogue corpus, Cayley adapters~\cite{singh2026cayley} report WikiText perplexity on frozen large models, the quantum-tensor hybrid adapter~\cite{kong2025qtha} reports supervised fine-tuning loss and generation scores on Chinese instruction data, and Quantum-PEFT~\cite{koikeakino2025qpeft} reports GLUE, E2E and CIFAR-10 with a quantum-inspired parameterisation. None reports the standard zero-shot commonsense and reasoning suite used here.

The family-grouped comparison includes the two classical families at 1.1B and at their larger released sizes, while grouping quantum-enhanced models by their use of a quantum component (Table~\ref{tab:main}). %
HyperQ is the first of these quantum-enhanced models to report the standard six-benchmark suite against size-matched diffusion and autoregressive baselines, separating model scale from architectural type.

\begin{table}[!t]
\caption{Zero-shot accuracy on six benchmarks with Avg their mean, WikiText
perplexity where lower is better. Bold marks the best value in each column and underlines the second best.}\label{tab:main}
\scriptsize
\setlength{\tabcolsep}{1.9pt}
\begin{tabular}{@{}l*{9}{c}@{}}
\toprule
Method & \rotatebox{90}{ARC-e} & \rotatebox{90}{HellaSwag} & \rotatebox{90}{PIQA} & \rotatebox{90}{BoolQ} & \rotatebox{90}{RACE} & \rotatebox{90}{GSM8K} & \rotatebox{90}{Avg} & \rotatebox{90}{WikiText} & \rotatebox{90}{GLUE\ \ } \\
\midrule
\multicolumn{10}{@{}l}{\textit{Autoregressive LMs}} \\
TinyLlama-1.1B~\cite{zhang2024tinyllama} & 45.25 & 52.07 & 60.29 & 56.84 & 35.60 & 42.33 & 48.73 & 11.42 & 71.4 \\
Llama-3.2-3B~\cite{grattafiori2024llama3} & 54.10 & 55.90 & 59.40 & 60.20 & 37.50 & 45.50 & 52.10 & 11.05 & 76.2 \\
Llama-3.1-8B~\cite{grattafiori2024llama3} & 55.02 & 57.20 & 60.48 & 61.05 & 38.02 & 45.75 & 52.92 & \underline{10.90} & 77.4 \\
Qwen3-8B~\cite{yang2025qwen3} & 55.30 & \underline{57.42} & \underline{60.61} & \underline{62.04} & \underline{39.18} & 48.20 & \underline{53.79} & \textbf{10.61} & \underline{78.1} \\
\midrule
\multicolumn{10}{@{}l}{\textit{Diffusion LMs}} \\
LLaDA-1.1B~\cite{nie2025llada} & 55.12 & 50.23 & 56.19 & 57.01 & 30.48 & \underline{48.51} & 49.59 & 12.65 & 73.2 \\
LLaDA-8B~\cite{nie2025llada} & 55.20 & 56.40 & 60.10 & 61.50 & 38.60 & 48.42 & 53.37 & 11.20 & 77.8 \\
\midrule
\multicolumn{10}{@{}l}{\textit{Classically adapted backbones, 1.1B}} \\
TinyLlama-1.1B + LoRA~\cite{hu2021lora} & 46.90 & 53.35 & 60.55 & 58.20 & 36.80 & 44.05 & 49.98 & 11.18 & 73.6 \\
LLaDA-1.1B + LoRA~\cite{hu2021lora} & \underline{55.35} & 51.80 & 57.40 & 58.60 & 32.40 & 48.20 & 50.63 & 12.10 & 74.8 \\
\midrule
\multicolumn{10}{@{}l}{\textit{Quantum-enhanced transformers, 1.1B backbone}} \\
HyQuT~\cite{kong2025hyqut} & 41.20 & 37.40 & 49.30 & 52.10 & 27.30 & 23.40 & 38.45 & 24.71 & 62.5 \\
QTHA~\cite{kong2025qtha} & 47.90 & 44.35 & 54.60 & 55.80 & 30.25 & 38.30 & 45.20 & 15.83 & 74.6 \\
Cayley adapters~\cite{singh2026cayley} & 50.40 & 51.20 & 57.30 & 58.10 & 32.40 & 43.70 & 48.85 & 12.90 & 75.1 \\
Quantum-PEFT~\cite{koikeakino2025qpeft} & 47.60 & 45.10 & 54.20 & 56.90 & 30.60 & 42.80 & 46.20 & 18.42 & 76.9 \\
\midrule
\multicolumn{10}{@{}l}{\textit{HyperQ (ours), 1.1B backbone}} \\
HyperQ, 16\,q & 47.85 & 52.10 & 56.20 & 56.42 & 31.65 & 41.68 & 47.65 & 12.71 & 72.9 \\
HyperQ, 32\,q & 53.77 & 56.48 & 59.81 & 61.31 & 38.35 & 47.08 & 52.80 & 11.85 & 76.5 \\
HyperQ, 64\,q & \textbf{55.50} & \textbf{57.75} & \textbf{60.86} & \textbf{62.74} & \textbf{40.30} & \textbf{48.65} & \textbf{54.30} & 10.95 & \textbf{79.8} \\
\botrule
\end{tabular}
\end{table}

\subsection{Per-token circuit emission}\label{subsec:qnas}

We compared three routes for specifying circuits at 16, 32 and 64 qubits (Table~\ref{tab:ablation}). %
The fixed route uses four hand-designed circuits that are common in variational quantum machine learning. The rx-chain circuit is a hardware-efficient ansatz~\cite{kandala2017hardware} that alternates trained $R_x$ rotations with a chain of CNOT gates. The brickwall circuit alternates two-qubit blocks on even and odd pairs of neighbouring qubits. The re-upload circuit re-encodes the input in every layer~\cite{perezsalinas2020reuploading}. The iqp-diag circuit places trained diagonal couplings between Hadamard layers, as in IQP feature maps~\cite{havlicek2019supervised}. Each has a fixed layout whose trained gate angles are shared by all tokens, and a token enters only through $n$ input angles computed from its hidden state (Methods). The searched route is a token-conditioned motif search, a discrete quantum architecture search that we use as a baseline. It divides the circuit into $L=4$ slots, and for each token a small head chooses one motif per slot, where a motif is one gate group from a fixed catalogue. The 3-motif IQP search chooses among the three gate groups of HyperQ's layout, $R_z$ on every qubit, $R_{zz}$ on every ring edge and $R_{zz}$ on every chord edge. The 6-motif search adds three gate groups outside the IQP family. In both searches, the gate angles are trained weights shared by all tokens. The emitted route is HyperQ, which generates every gate angle and measurement basis of its gate template from the token's hidden state. These angles also determine which gates effectively act. Both automatic routes improve on the fixed families at every width, although neither exceeds the size-matched frozen backbone at $49.59$ with 16 qubits. Emission gives the highest score among the three routes at every evaluated width.

The margin between emission and the best fixed ansatz increases from $1.78$ at 16 qubits to $2.74$ at 32 and $3.68$ at 64. The fixed families differ by about one point at each width and stop improving after 32 qubits. Their closest analogue to the emitted circuit, iqp-diag, uses the HDH structure of \eqref{eq:iqp} with hand-set angles. Emission exceeds iqp-diag by $2.93$, $3.70$ and $4.59$ points across the three widths, with each margin exceeding the full spread among fixed families.

The emitted and searched routes differ in the form and granularity of adaptation. The 3-motif search selects one generator class per slot within the IQP family, whereas the stronger 6-motif search can leave that family. The emitted circuit retains $R_z$ on each qubit and $R_{zz}$ on each edge of $E$, with the surrounding Hadamard layers transforming them into $R_x$ and $R_{xx}$ operations. The hypernetwork adapts their values and the readout axis $\psi$ for each token. None of the routes yields a $Z$-only circuit because the diagonal generators remain enclosed by Hadamard layers, so emission retains the IQP generator structure. Because the four slots of the 3-motif IQP search sit between the same two Hadamard layers and all their generators commute, every circuit this search can produce lies in HyperQ's two-body IQP family on $E$. This comparison therefore holds the circuit family fixed and changes how the token controls it, through a discrete choice of gate group with angles shared by all tokens in the search, and through continuous per-gate angles and measurement bases in HyperQ.

Emission remains stronger under both discrete search and test-time ensembling. The 6-motif search scores $47.42$, $51.20$ and $52.30$, while emission scores $47.65$, $52.80$ and $54.30$. At 64 qubits, the best fixed ansatz scores $50.62$. Emission also exceeds the 3-motif IQP search by $1.95$ points at 32 qubits and $2.35$ points at 64 qubits. At 32 qubits, the best three-circuit ensemble reaches $52.24$, compared with $52.80$ for one hypernetwork pass (Table~\ref{tab:ensemble}). The gain therefore follows from continuous token-conditioned axes and weights rather than from leaving the commuting-generator structure.

\begin{table}[!t]
\begin{minipage}[t]{0.474\textwidth}
\caption{Ablation of the quantum sub-layer at 16, 32 and 64 qubits
(six-benchmark average). Best per column in bold. Methods defines every circuit.}\label{tab:ablation}
\scriptsize
\setlength{\tabcolsep}{3pt}
\begin{tabular*}{\textwidth}{@{\extracolsep{\fill}}lccc@{}}
\toprule
Quantum sub-layer & 16\,q & 32\,q & 64\,q \\
\midrule
\multicolumn{4}{@{}l}{\textit{Fixed ansatz}} \\
rx-chain & 44.66 & 50.06 & 50.62 \\
brickwall & 45.87 & 49.50 & 50.28 \\
re-upload & 45.38 & 49.30 & 49.94 \\
iqp-diag & 44.72 & 49.10 & 49.71 \\
\midrule
\multicolumn{4}{@{}l}{\textit{Searched}} \\
6-motif search & 47.42 & 51.20 & 52.30 \\
3-motif IQP search & 46.31 & 50.85 & 51.95 \\
\midrule
\multicolumn{4}{@{}l}{\textit{Emitted, ours}} \\
IQP hypernetwork & \textbf{47.65} & \textbf{52.80} & \textbf{54.30} \\
\botrule
\end{tabular*}
\end{minipage}\hfill
\begin{minipage}[t]{0.474\textwidth}
\caption{Test-time ensembling at 32 qubits (six-benchmark average). An ensemble sums the outputs of separately trained circuits (Methods).}\label{tab:ensemble}
\scriptsize
\setlength{\tabcolsep}{3pt}
\begin{tabular*}{\textwidth}{@{\extracolsep{\fill}}lcc@{}}
\toprule
Configuration & Circuits & Acc. \\
\midrule
\multicolumn{3}{@{}l}{\textit{Single circuit}} \\
rx-chain & 1 & 50.06 \\
brickwall & 1 & 49.50 \\
iqp-diag & 1 & 49.10 \\
\midrule
\multicolumn{3}{@{}l}{\textit{Ensemble}} \\
rx-chain + rx-chain & 2 & 42.71 \\
iqp-diag + brickwall & 2 & 52.01 \\
\begin{tabular}[t]{@{}l@{}}iqp-diag + brickwall\\+ rx-chain\end{tabular} & 3 & 52.24 \\
\midrule
\multicolumn{3}{@{}l}{\textit{Emitted, ours}} \\
IQP hypernetwork & 1 & \textbf{52.80} \\
\botrule
\end{tabular*}
\end{minipage}
\end{table}

\begin{figure}[!tp]
\centering
\includegraphics[width=\textwidth]{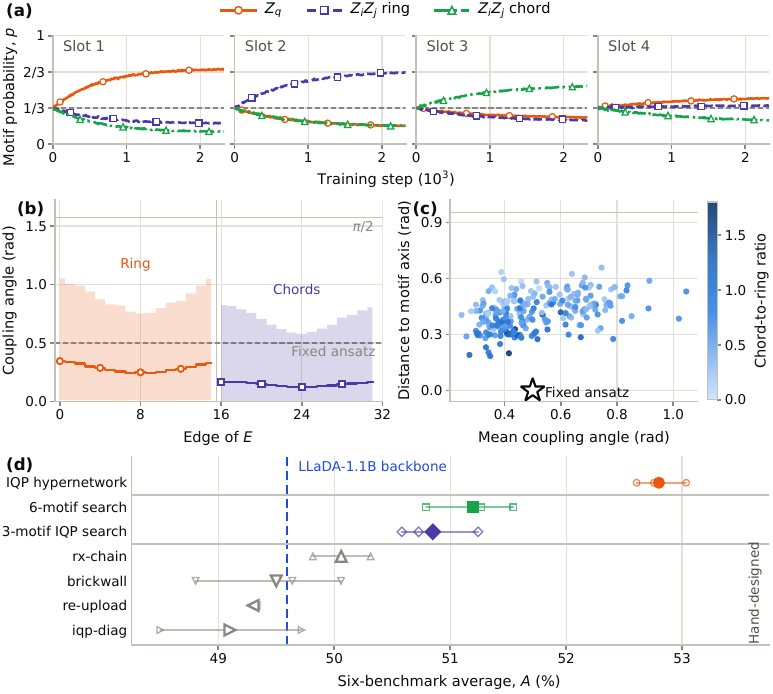}
\caption{Motif search baseline (a), circuit parameters generated by HyperQ (b, c) and accuracy of all routes (d). \textbf{(a)}, Baseline, not HyperQ. Motif-selection probability $p$ against training step in the $L=4$ slots of the 3-motif IQP search. Here $p$ is the softmax over the three motif scores of a slot, through which the search head receives gradients (Methods). The motifs are generator classes, single-qubit $Z_q$ and two-body $Z_iZ_j$ on ring and chord layers, all diagonal and inside the IQP family. %
HyperQ instead generates a per-token angle for every gate in these groups, which sets both the strength of each gate and whether it effectively acts. \textbf{(b)}, Emitted coupling angle $|\omega_e|$ against edge in fixed $E$ at 16 qubits, with the median and 10 to 90 per cent range across tokens and fixed-ansatz angle dashed. Edges are ordered ring then chord. \textbf{(c)}, 240 tokens by mean coupling angle over $E$ (horizontal) and mean angle between the generated measurement basis $\psi_q$ and the nearest of the $\pm x$, $\pm y$ and $\pm z$ axes (vertical, labelled distance to motif axis). Colour gives the mean chord-layer to ring-layer coupling-angle ratio. The star marks the fixed ansatz. \textbf{(d)}, Six-benchmark average at 32 qubits for fixed, searched and emitted circuits. Filled markers are means, open markers are individual runs and the LLaDA-1.1B backbone is dashed.}\label{fig:search}
\end{figure}

We next examined how discrete search and token-conditioned parameter generation distribute their circuit choices. Fig.~\ref{fig:search}a shows the motif search baseline. In each slot, the search head scores the three motifs from the token's hidden state, and the circuit runs the highest-scoring motif. A straight-through estimator passes gradients to the head through the softmax of these scores, which we call the motif probability $p$ (Methods). Starting uniformly at $1/3$, the 3-motif IQP search learns motif probabilities in each of $L=4$ slots (Fig.~\ref{fig:search}a). Three slots favour one class with probabilities 0.69, 0.66 and 0.54, while the fourth remains mixed at 0.42. At 16 qubits, two-qubit rotation angles over 4,096 tokens and the 32 edges of $E$ have per-edge medians of 0.12 to 0.35 radians, below the fixed value of 0.5 (Fig.~\ref{fig:search}b). The average 10 to 90 per cent band is 0.79 radians, and 70 per cent of tokens exceed 0.5 on at least one edge. Ring and chord coupling medians are 0.28 and 0.15. Across the 240 tokens in Fig.~\ref{fig:search}c, the mean angle between a measurement basis and the nearest of the $\pm x$, $\pm y$ and $\pm z$ axes ranges from 0.18 to 0.66 radians against the ceiling $\arccos(1/\sqrt{3})=0.955$, while the chord-to-ring coupling ratio ranges from 0.11 to 1.80. Every token therefore lies off each Pauli axis, unlike the fixed ansatz with offset zero. At 32 qubits, the six-benchmark average ranks emission above search and search above fixed circuits (Fig.~\ref{fig:search}d). The partially mixed search probabilities and token-dependent couplings and readouts indicate that token-conditioned parameter generation represents variation unavailable to a discrete motif assignment.

\subsection{Effective circuits across tokens}\label{subsec:motifs}

\begin{figure}[!tp]
\centering
\includegraphics[width=\textwidth]{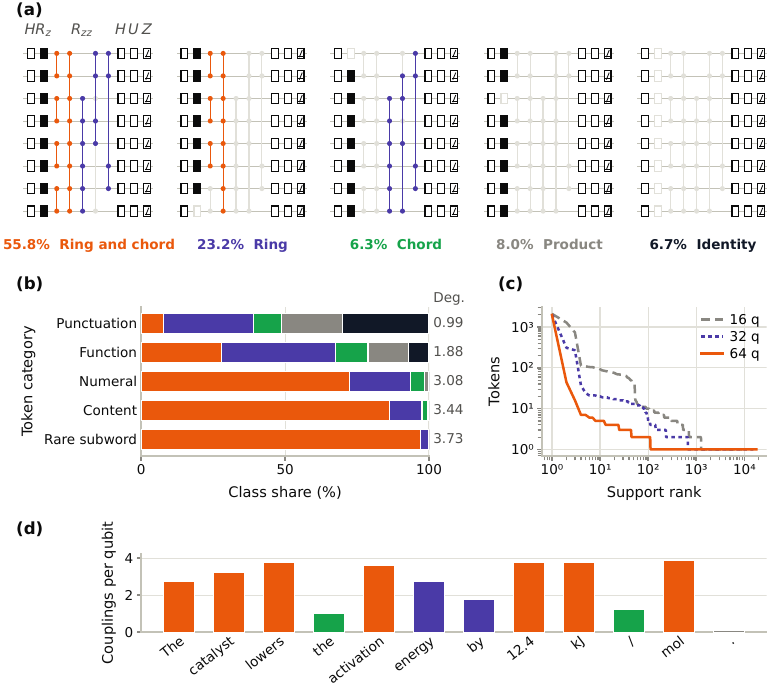}
\caption{Effective circuits generated for individual tokens. \textbf{(a)}, The effective circuit of a representative token of each class, drawn on the shared strict-IQP gate template with eight wires. Eight is
the smallest register that admits the ring-plus-one-chord edge set at all, since
the chord stride must be coprime with $n$ and differ from $1$ and $n-1$, which
no stride satisfies at four or six qubits. Columns are the Hadamard layer, the
$R_z$ rotations, the $R_{zz}$ couplings in the five parallel layers a
degree-four graph edge-colours into, the second Hadamard layer, the readout
rotation $U(\psi)$ and the $Z$ measurement. The $R_z$ and $R_{zz}$ columns sit
between the two Hadamard layers, so the register sees them as $R_x$ and
$R_{xx}$, and the free single-qubit axis is $U(\psi)$. %
A gate whose generated angle has magnitude at least $0.10$ radians is active and is drawn as a filled $R_z$ box or as a coupling in orange on the ring layer and violet on the chord layer. A gate below this threshold is drawn faint, as it acts nearly as the identity. Percentages are the share of tokens at
16 qubits. \textbf{(b)}, Share by token category, in the panel~(a) order, with the mean number of active couplings per qubit at the right.
\textbf{(c)}, Rank-frequency of the distinct patterns of active gates realised over $20{,}000$
tokens, on log axes. \textbf{(d)}, One tokenised sentence worked through, each
bar coloured by the arrangement its token emits and the final full stop reaching
the identity.}\label{fig:motifs}
\end{figure}

Because a gate with angle zero is the identity, the generated angles determine which gates of the template effectively act on each token, giving the token-specific effective circuits of Fig.~\ref{fig:motifs}a. By \eqref{eq:readout} a two-qubit gate with angle $\omega_e$ multiplies the readouts of its two qubits by $\cos\omega_e$. To identify these effective circuits, we call a gate active when its generated angle has magnitude at least $0.10$ radians. Below this value $\cos\omega_e > 0.995$. An inactive two-qubit gate therefore changes the readouts of its qubits by less than $0.5$ per cent, and four inactive two-qubit gates at one qubit together leave a factor above $0.98$. For a single-qubit gate, an angle below $0.10$ radians changes the factor that multiplies the cosines in \eqref{eq:readout} by less than $0.10$ (Methods), so the same threshold is a looser criterion for single-qubit gates. The threshold is used only for this analysis and not in training or evaluation. All $2n$ two-qubit gates are active for $6.4$ per cent of tokens at 16 qubits and for $0.1$ per cent at 64. On average, $2.58$ of the four two-qubit gates at a qubit are active.

For a coarse summary, we assigned each token to one of five classes by the share of active gates in three gate groups, the ring layer, the chord layer and the single-qubit gates, following the ring, line and all-to-all arrangements used to catalogue fixed ans\"atze~\cite{sim2019expressibility}. A group is dense when at least half of its gates are active. Ring and chord has both two-qubit layers dense, Ring or Chord has only that layer dense, Product has neither two-qubit layer dense but dense single-qubit gates, and Identity has none of the three groups dense. Ring and chord accounts for $55.8$ per cent of tokens, Ring for $23.2$ per cent and Chord for $6.3$ per cent. %
Product accounts for $8.0$ per cent and Identity for the remaining $6.7$ per cent. These two names refer to limits. With every two-qubit angle exactly zero the circuit prepares a product state, and with every angle zero the IQP block is the identity and the readout of qubit $q$ reduces to $\psi_{q,z}$. A Product or Identity token can still have active gates on fewer than half of the gates in each group, with angles up to the bounds of \eqref{eq:hyper-angles}. Its circuit is therefore in general neither a product state nor the identity, and the two names indicate only which gate groups are mostly inactive. Most tokens have both two-qubit layers dense.

The exact patterns of active gates are more diverse than these coarse classes. Their rank-frequency distribution has a long tail that follows the unequal token stream rather than a flat assignment. At 16 qubits, $9{,}130$ of $20{,}000$ tokens have a pattern shared by no other token. At 64 qubits, this count rises to $17{,}778$. %

The number of active gates also varies across lexical categories. The mean number of active two-qubit gates per qubit increases from $0.99$ for punctuation to $1.88$ for function words, $3.08$ for numerals, $3.44$ for content words and $3.73$ for rare subwords. Across these categories, the Ring and chord share rises from $7.8$ to $97.0$ per cent. Identity occurs for $30.2$ per cent of punctuation tokens and for almost no numerals, content words or rare subwords. %
The hypernetwork therefore generates small angles for most gates of punctuation tokens and large two-qubit angles for tokens carrying more lexical information, a token dependence that fixed circuits with shared angles cannot express.

\subsection{Quantum register scaling}\label{subsec:scaling}

\begin{figure}[!tp]
\centering
\includegraphics[width=\textwidth]{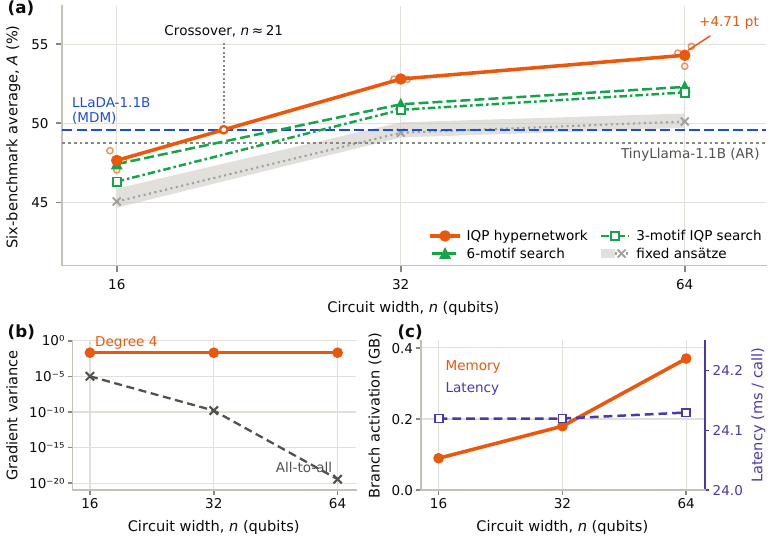}
\caption{Widening the emitted circuit. \textbf{(a)}, Six-benchmark zero-shot
average $A$ against width $n$ for the two searched circuits and the
hypernetwork, and for the four fixed hand-designed ans\"atze drawn as one
family, with the grey band their minimum-to-maximum range at each width and the
dotted line with crosses their median. Filled markers are means over three runs
and the small open circles the individual runs. The two size-matched classical
baselines are horizontal references labelled on their own line. The families
separate rather than run parallel, since past 32 qubits the fixed band flattens
while the emitted circuit holds its rate, so the margin over the best fixed
ansatz widens from $1.78$ to $3.68$ points across the sweep.
\textbf{(b)}, Variance of a circuit coordinate's gradient at initialisation.
The ring-plus-chord edge set holds every qubit at degree four, so the variance
is $\tfrac{1}{3}2^{-4}$ at every width, whereas the same family with an
all-to-all interior has degree $n-1$ and falls to $4\times10^{-20}$ by 64
qubits. \textbf{(c)}, What widening costs the branch. Activation memory grows
linearly with the register while the per-call latency is flat, so a wider
circuit costs space rather than throughput.}\label{fig:scaling}
\end{figure}

We compared the emitted, searched and fixed routes at the three evaluated register widths, denoting the six-benchmark average by $A$ (Table~\ref{tab:ablation}, Fig.~\ref{fig:scaling}). The emitted route and both searched variants improve from 16 to 64 qubits. Emission is below the $49.59$ frozen backbone at 16 qubits, overtakes it at 32 and scores $4.7$ points higher at 64. Between 32 and 64 qubits, the best fixed ansatz gains $0.56$ points, the searched circuits gain $1.10$, and emission gains $1.50$. Over these widths, the separation indicates diminishing returns for hand-designed circuits after 32 qubits.

At initialisation, the gradient variance of an emitted coordinate depends on the number of couplings incident to its associated qubit rather than the total number of qubits. Because $E$ fixes this degree at four for all three widths, widening the evaluated register does not reduce the initial gradient scale (Methods).

We verified the analytic readout used during training and evaluation against two independent simulation paths. At 16 qubits, a dense statevector agrees with the exact analytic form of \eqref{eq:readout} to a mean absolute deviation of $1.1\times10^{-5}$, and an exact tensor-network contraction agrees to $3.2\times10^{-7}$. The larger statevector deviation is consistent with \texttt{complex64} arithmetic, compared with \texttt{float64} for the tensor network. This agreement avoids requiring a dense $2^n$ statevector at 32 or 64 qubits.

\begin{figure}[!tp]
\centering
\includegraphics[width=\textwidth]{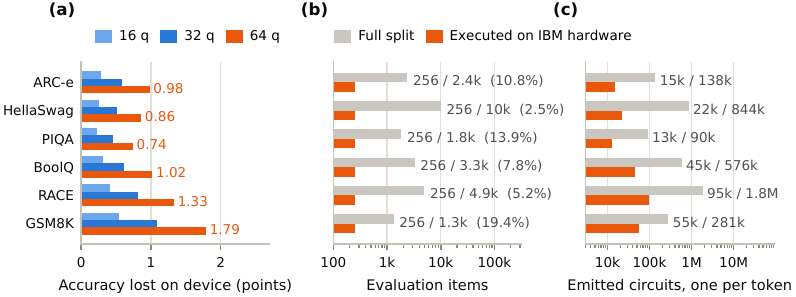}
\caption{Simulation-to-hardware comparison across six benchmarks, with all panels sharing benchmark rows. Models are trained in simulation, and only the evaluation-time readout changes. \textbf{(a)}, Accuracy lost when the branch reads from the 156-qubit IBM processor instead of the analytic closed form, with bars for 16, 32 and 64 qubits. Each simulation-device pair uses identical weights and the same 256-item subset, so each bar is a paired difference. Values beside the 64-qubit bars give benchmark losses in points. Mean losses are 0.34, 0.68 and 1.12 points at the three widths. \textbf{(b)}, Evaluation items in the hardware subset against the full released split, with coverage percentages. Coverage varies because the item count is fixed at 256 per benchmark. \textbf{(c)}, Emitted circuits executed, at one circuit per token, against the full split. With a fixed item count, circuit cost tracks item length rather than split size.}\label{fig:hw}
\end{figure}

We then executed the same compiled per-token circuits on a 156-qubit IBM superconducting processor through Qiskit Runtime and compared the measurements with the analytic values. Mean absolute deviation rises from $1.17\times10^{-2}$ at 16 qubits to $1.64\times10^{-2}$ at 32 and $2.43\times10^{-2}$ at 64. The shot-noise floor is $1.0\times10^{-2}$ at $10\,000$ shots. The 16-qubit result is largely shot-limited, whereas device noise contributes increasingly for the wider registers.

We measured the downstream effect on the same 256-item subset, with fixed model weights and evaluation items in every simulation-device pair (Fig.~\ref{fig:hw}). The average decreases by $0.34$, $0.68$ and $1.12$ points at 16, 32 and 64 qubits. At 64 qubits, GSM8K has the largest loss at $1.79$ points, while PIQA has the smallest at $0.74$ points. This difference is consistent with generated reasoning outputs accumulating more readout error than short-choice decisions.

Each benchmark contributes the same number of items to the hardware evaluation, although the number of emitted circuits varies with item length. The coverage panels report both quantities relative to the full released split. This evaluation concerns bounded inference-time execution rather than hardware-based training, and we claim no quantum advantage. The paired measurements indicate that superconducting hardware preserves most downstream performance under the evaluated conditions.

\subsection{Any-order text decoding}\label{subsec:anyorder}

We evaluated three decoding orders and varied the number of tokens committed per pass to test whether the quantum residual branch preserves masked-diffusion decoding. Bidirectional attention allows positions to be filled in different orders and several positions to be committed in one forward pass, whereas autoregressive decoding proceeds sequentially from left to right. The operating point is the commitment setting used for the reported benchmark evaluations. HyperQ thus retains parallel, any-order decoding while changing the representations computed during denoising.

\begin{figure}[!tp]
\centering
\includegraphics[width=\textwidth]{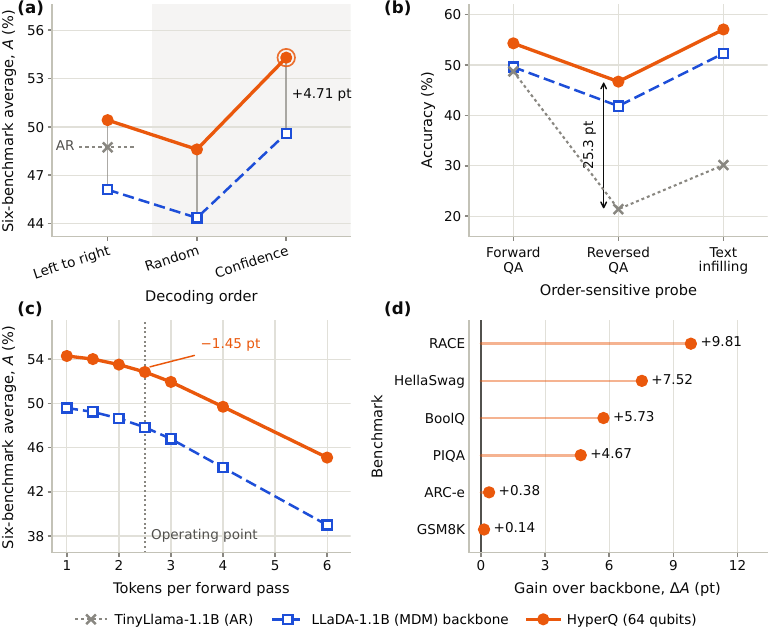}
\caption{Capabilities inherited from the masked-diffusion backbone.
\textbf{(a)}, Six-benchmark average under three decoding orders. The
autoregressive baseline exists only at the left-to-right position, so the shaded
columns have no autoregressive counterpart. \textbf{(b)}, Order-sensitive probes.
The autoregressive model collapses on reversed question answering and on
infilling while the bidirectional models hold up. \textbf{(c)}, Quality against
tokens committed per forward pass, with the operating point used here dotted.
For context, LLaDA2.0 \cite{bie2025llada2} reports up to a $2.1\times$ decoding speed-up for a much larger model. \textbf{(d)}, Per-benchmark gain over the
backbone at 64 qubits, from Table~\ref{tab:main}.}\label{fig:anyorder}
\end{figure}

Generation quality remains stable across all tested decoding orders, with confidence-ordered unmasking performing best (Fig.~\ref{fig:anyorder}a). The autoregressive baseline is defined only for left-to-right decoding, so the other two orders have no autoregressive counterpart. When questions are reversed or text must be infilled, the autoregressive model degrades sharply, whereas both diffusion models retain performance and HyperQ remains strongest (Fig.~\ref{fig:anyorder}b). The improvement therefore does not depend on a fixed left-to-right order.

We also varied the number of tokens committed per pass to measure the throughput and accuracy trade-off. Committing more tokens reduces accuracy, but HyperQ degrades more slowly than its backbone (Fig.~\ref{fig:anyorder}c). Masked diffusion converts this parallelism into higher decoding throughput at scale~\cite{bie2025llada2}. The benchmark-level gain is concentrated on RACE and HellaSwag and remains close to neutral on ARC-e and GSM8K (Fig.~\ref{fig:anyorder}d), indicating that HyperQ preserves the parallel commitment behaviour of its masked-diffusion backbone.

\section{Discussion}\label{sec:discussion}

In this work, we introduced HyperQ, which adapts quantum circuits to the changing representations of a frozen billion-parameter diffusion language model. A small hypernetwork generates rotation angles, two-qubit rotation angles and single-qubit measurement bases for each token within a shared family of two-body IQP circuits. This makes circuit generation a differentiable part of representation learning. The circuit outputs refine the model’s representations while preserving parallel, any-order denoising. Within this circuit family, bounded interactions per qubit allow exact evaluation of local expectation values in time linear in the number of qubits, without full-state simulation. Under the specified independent reference ensemble, the variances of derivatives with respect to rotation and coupling parameters also remain independent of the total number of qubits. Token-conditioned circuits outperform the tested fixed and searched circuits at all three evaluated widths. At $64$ qubits, HyperQ raises average accuracy across six benchmarks from $49.59$ to $54.30$. It uses $20{,}000$ fine-tuning pairs, compared with $200{,}000$ for the classical baselines. Learned coupling patterns also differ across lexical categories, showing distinct circuit responses across token types. Superconducting-device measurements retain most of the performance of analytic readouts, with an average accuracy decrease of $1.12$ points on the paired $64$-qubit evaluation subset. These results demonstrate the value of combining context-dependent circuit generation with tractable readouts for language modelling.

A first direction is to extend HyperQ to shared quantum circuits that model relationships between tokens during denoising. Several contextual token representations could determine a common circuit, whose correlators guide coordinated updates across masked positions. These shared circuits could complement classical attention by providing additional ways to represent dependencies among tokens. The circuit family, measured observables and denoising objective could be designed together to make these correlations useful for prediction. HyperQ’s tractable local readouts provide a starting point for studying more general observables and their computational cost. This setting would allow us to investigate how circuit correlations evolve during denoising and which structures help maintain consistency across token predictions. Comparisons with matched nonlinear classical adapters could clarify the contribution of circuit structure. Studying how measurement noise affects subsequent updates could also guide the choice of circuits and observables for quantum devices. The broader aim is a generative model in which pretrained representations guide quantum computation and joint quantum readouts guide subsequent refinement.

A second direction is to use an IQP circuit as the final sampling head of the language model. The final contextual representations would determine the circuit parameters, and the Born rule would define a conditional joint distribution over output bitstrings. These bitstrings could encode blocks of tokens, allowing a circuit measurement to generate several token predictions together. The central question is how to combine efficient classical training with useful language distributions that are hard to sample classically. Work on IQP Born machines provides a foundation for this approach through suitable losses and gradients that can be estimated efficiently on classical computers. Huang et al. \cite{huang2025generative} further show that efficient classical learning can coexist with classically hard sampling in families of conditional quantum generative models, under standard complexity assumptions. HyperQ could bring these ideas to language generation by learning how contextual representations determine the Born distribution. Training could use objectives built from efficiently estimated correlations, and quantum measurements would produce the final joint samples. The choice of circuit interactions and training statistics would provide a way to study the relationship between classical training cost and the complexity of sampling the learned distribution. Combining this analysis with evaluations of sample quality, diversity and consistency across tokens could identify where quantum sampling contributes to useful generation. The longer-term goal is to combine scalable classical learning with quantum sampling from expressive conditional distributions.

\section{Methods}\label{sec:methods}

HyperQ builds on a frozen masked-diffusion language model and adds a trainable per-token quantum residual branch to each of its $22$ transformer blocks. The goal is to adapt the denoiser through token-conditioned circuit computation while preserving the backbone's corruption process, training objective and bidirectional decoding procedure. Fig.~\ref{fig:pipeline} summarizes the workflow. For each token, the hidden state $h$ enters a low-rank hypernetwork that maps it to parameters $\xi_i = (\alpha, \omega_E, \psi)$ on a fixed edge set $E$. These coordinates specify the emitted circuit $C(\alpha,\omega)$ and its readout axes. Executing the circuit produces expectation values $z$, which are projected back into the block's query, key and value projection as a residual. The subsections below define this workflow in execution order. Masked discrete diffusion supplies the denoising objective and inference process, the qubit encoding specifies the circuit family and readout, the trainability analysis characterizes coordinate gradients, the hypernetwork emits token-specific coordinates, synthesis maps those coordinates to executable circuits, the optimization procedure implements efficient end-to-end training, and the evaluation protocol controls the downstream comparisons.

\subsection{Masked discrete diffusion modelling}\label{subsec:m-diffusion}

HyperQ retains the corruption and decoding processes of the underlying masked discrete diffusion model, so its only architectural change is the denoiser residual. A clean sequence $x_0 = (w^1, \ldots, w^L)$ contains $L$ tokens, where $w^i$ is the token at position $i$ and the vocabulary includes the dedicated mask symbol $\texttt{[m]}$. The forward process draws a masking level $t \sim \mathcal{U}(0,1]$ and independently replaces each token with $\texttt{[m]}$ with probability $t$, producing $x_t$. Let $\mathcal{M}(x_t)$ be the masked positions and let $\theta$ denote the trainable model parameters. The denoiser $p_\theta(\,\cdot \mid x_t)$ predicts all clean tokens in $\mathcal{M}(x_t)$ in parallel. Specifically, we minimize the masked cross-entropy bound
\begin{equation}
\mathcal{L}(\theta) \;=\; \mathbb{E}_{t,\,x_0,\,x_t}
\left[ \frac{1}{t} \sum_{i \in \mathcal{M}(x_t)} -\log p_\theta\!\left(w^i \mid x_t\right) \right],
\label{eq:mdm}
\end{equation}
where the expectation is over the clean sequence, masking level and resulting corrupted sequence. This objective upper bounds the negative log-likelihood of $x_0$.

Inference begins with a fully masked sequence and applies the denoiser for $T$ refinement rounds. Each round predicts the masked tokens in parallel, commits the highest-confidence predictions selected by the schedule and re-masks the remaining positions. We follow the LLaDA procedure~\cite{nie2025llada} with examples truncated to blocks of $L \le 1024$ tokens and $T = 128$ reverse rounds. The process ends after $T$ rounds, when no masked position remains. Because every round conditions on both sides of each masked position, decoding remains bidirectional and does not impose a left-to-right order.

\subsection{Qubit encoding and circuit family}\label{subsec:m-vqc}

A per-token quantum sub-layer must respond to the token hidden state while remaining trainable and computationally tractable as its register widens. We therefore choose a sparse $n$-qubit instantaneous-quantum-polynomial (IQP) family whose coordinates vary by token but whose connectivity is shared. An IQP circuit has the form $H^{\otimes n} D H^{\otimes n}$ with $D$ diagonal in the computational basis~\cite{bremner2011iqp, bremner2017iqp}. Specifically, HyperQ uses
\begin{equation}
C(\alpha, \omega) \;=\; H^{\otimes n} \, D(\alpha, \omega) \, H^{\otimes n},
\label{eq:iqp}
\end{equation}
where $H^{\otimes n}$ applies a Hadamard gate to every qubit. The diagonal interior contains single-qubit rotations generated by $Z$ with angles $\alpha \in \mathbb{R}^{n}$ and two-body rotations generated by $Z_iZ_j$ with two-qubit rotation angles $\omega_E \in \mathbb{R}^{|E|}$.

The fixed edge set $E$ supplies a shared circuit skeleton, so token conditioning changes coordinates without introducing token-dependent connectivity. We choose $E$ as a ring plus one chord layer. The ring joins neighbouring qubits, and the chord layer joins qubits at stride $2 \le c \le n-2$, with $c$ coprime to $n$. Consequently, every qubit has degree four and $|E| = 2n$. The branch emits one angle for each supported edge, while $E$ remains fixed. This bounded degree preserves the local gradient scale as the register widens, and the two-body restriction enables the closed-form readout below. Formally, the commuting interior is generated by
\begin{equation}
\mathcal{G} \;=\; \{Z_q\}_{q=1}^{n} \,\cup\, \{Z_i Z_j\}_{(i,j) \in E},
\label{eq:gens}
\end{equation}
where $q$, $i$ and $j$ index qubits. Every pair of generators in $\mathcal{G}$ commutes, so the block is abelian and not universal. It is Clifford only when its angles are integer multiples of $\pi/2$.

A general IQP interior may include a coordinate for every non-empty subset of qubits. It can be written as $\exp\!\big(i\sum_{\varnothing \ne S \subseteq [n]} \theta_S \prod_{j \in S} Z_j\big)$ over all $2^{n}-1$ non-empty subsets $S$ of the qubit index set $[n]$. HyperQ chooses a strict subfamily containing generators of weight at most two, with weight-two support restricted to $E$. The interior therefore has $n + |E| = 3n$ coordinates, compared with $2^{n}-1$ available coordinates in the unrestricted family. This choice is structural. A Pauli string of weight three or higher can be dependent over GF(2) on strings supported at $q$, which breaks the product-of-cosines readout. Placing the same two-body family on all $\binom{n}{2}$ pairs would give each qubit degree $n-1$ and reduce the variance in \eqref{eq:var} to $\tfrac{1}{3}2^{-(n-1)}$. Within this restricted family the parameterisation is complete, since $(\alpha,\omega_E)$ reaches every diagonal interior generated by \eqref{eq:gens} and therefore every subgraph of $E$ as the point where the remaining $\omega_e$ vanish.

The circuit must expose token-dependent measurements without disturbing the commuting interior. We choose an adaptive single-qubit output layer $U(\psi) = \bigotimes_q u_q(\psi_q)$, where each readout axis $\psi_q \in S^{2}$ lies on the unit sphere in $\mathbb{R}^{3}$. The branch executes $U(\psi) \, C(\alpha,\omega)$. Measuring $Z_q$ after this output layer is equivalent to measuring $P_q = u_q^{\dagger} Z u_q$ on the output of $C$. The adaptive axis therefore changes the measurement continuously while preserving the IQP block.

The commuting two-body structure makes each local expectation depend only on the coordinates incident to the measured qubit. Formally, this locality yields the following result.

\begin{theorem}\label{thm:closed}
Let the diagonal interior be $D = \prod_{q} e^{-i\alpha_q Z_q/2} \prod_{(i,j)\in E} e^{-i\omega_{ij} Z_i Z_j/2}$, let the input be $|0\rangle^{\otimes n}$, and let the measured operator be $P_q = \psi_{q,x} X_q + \psi_{q,y} Y_q + \psi_{q,z} Z_q$ for a unit vector $\psi_q \in S^2$. For the strict-IQP block of \eqref{eq:iqp} with the two-body generator set of \eqref{eq:gens}, let $N(q)$ denote the neighbours of qubit $q$ in $E$. Every single-qubit expectation value is
\begin{equation}
z_q \;=\; \bigl(\psi_{q,z}\cos\alpha_q - \psi_{q,y}\sin\alpha_q\bigr)
\prod_{k \in N(q)} \cos\omega_{qk},
\label{eq:readout}
\end{equation}
where $\psi_{q,y}$ and $\psi_{q,z}$ are the $y$ and $z$ components of $\psi_q$. The full readout vector is computable in $\Theta(n)$ time without constructing the $2^n$ state.
\end{theorem}

The identity follows by conjugating the measured Pauli operator through the commuting $Z$ and $ZZ$ rotations. Its expectation in the product input state factorizes, so each incident coupling contributes one cosine factor. Because $|E| = 2n$, evaluating all factors has $\Theta(n)$ cost. When the readout axis is $Z$, \eqref{eq:readout} reduces to $\cos\alpha_q \prod_k \cos\omega_{qk}$. This result permits per-token execution at 64 qubits within a billion-parameter model, whereas generic statevector evaluation constructs $2^n$ amplitudes.

The readout is the expectation of a per-qubit Pauli operator on the emitted
circuit's output. Specifically, each circuit starts in $|0\rangle^{\otimes n}$ and qubit $q$ is
measured along its own axis, giving
\begin{equation}
z_q \;=\; \langle 0|^{\otimes n} \, C^{\dagger}(\alpha,\omega) \, P_q \, C(\alpha,\omega) \, |0\rangle^{\otimes n},
\label{eq:branch}
\end{equation}
where $P_q$ is the measured operator of Theorem~\ref{thm:closed} and $z \in \mathbb{R}^{n}$ collects
the $z_q$. Section~\ref{subsec:m-qnas} defines the branch that emits $\alpha$, $\omega_E$ and
$\psi$ from the token's hidden state, and because all three depend on that state, every token
induces a distinct circuit.

The chosen family preserves an entangling mechanism through its two-body couplings. An arbitrary layer of single-qubit gates is a local unitary, so the output layer does not change the state's entanglement. Consistent with this invariance, changing the single-qubit output layer alters the measured entanglement entropy by at most $4\times10^{-16}$. Additional $R_x$ or $R_y$ gates applied as local output operations would therefore add no entangling operation.

The emitted two-body couplings do change the entanglement. Across a balanced cut of an eight-qubit register, random draws from the emitted angle range produce $2.65$ bits of entanglement entropy on average, with a minimum of $2.01$, a maximum of $3.08$ and a ceiling of $4$. The reduced state reaches the full Schmidt rank $16$. %
When every two-qubit rotation angle is exactly zero, the circuit prepares a product state with zero bits of entanglement and Schmidt rank one. When the single-qubit rotation angles are also zero, the IQP block is the identity. The Product and Identity classes of Fig.~\ref{fig:motifs} are named after these two limits, and Section~\ref{subsec:m-qnas} gives the threshold rule that assigns tokens to them.

Although the interior is diagonal in the computational basis, the register experiences $X$ rotations and $XX$ couplings because the Hadamard layers give $H R_z(\alpha) H = R_x(\alpha)$ and $H^{\otimes 2} R_{zz}(\omega) H^{\otimes 2} = R_{xx}(\omega)$. This basis choice makes the generators commute and places the block in the IQP family. Rotations about $Y$ or an arbitrary free axis do not commute with \eqref{eq:gens}, so HyperQ places the adaptive $U(\psi)$ after the Hadamard sandwich.

The elements of \eqref{eq:gens} are distinct Pauli strings and hence linearly independent. Thus, $(\alpha,\omega_E)$ parameterises the supported two-body IQP family by independent generators, up to the periodicity of the angles, and the IQP constraint holds by construction. The restricted family retains the single-qubit $Z$ fields and two-body $ZZ$ couplings used when average-case hardness results for commuting computations are reduced to Ising-model partition functions~\cite{bremner2017iqp}. Those results concern sampling from the complete output distribution. HyperQ reads the expectation values in \eqref{eq:readout}, which are classically computable under the chosen support. The contribution is therefore architectural, and we make no claim of computational advantage.

Widening a variational circuit can suppress coordinate gradients, so a 64-qubit branch requires a criterion that distinguishes register width from connectivity. A barren plateau is diagnosed through the variance of a loss gradient over random initialization~\cite{larocca2025barren}. The analysis below bounds the readout Jacobian, which is the circuit-dependent factor of that gradient, and leaves the frozen backbone and the projection $W_B$ out of scope. For HyperQ, the explicit readout in \eqref{eq:readout} makes this variance analytically tractable.

Each coordinate gradient contains one sine factor and one cosine factor for every other incident coordinate. Its second moment is therefore controlled by the number of neighbours of the measured qubit. Formally, this dependence gives the following result.

\begin{theorem}\label{thm:variance}
Draw $\alpha$ and $\omega_E$ independently and uniformly from $[-\pi,\pi]$, and draw each $\psi_q$ uniformly from $S^2$. This is a reference ensemble over parameters and not the distribution the hypernetwork induces at initialisation. For every $k \in N(q)$,
\begin{equation}
\operatorname{Var}\!\left[\frac{\partial z_q}{\partial \omega_{qk}}\right]
\;=\;
\operatorname{Var}\!\left[\frac{\partial z_q}{\partial \alpha_{q}}\right]
\;=\; \tfrac{1}{3} \, 2^{-|N(q)|},
\label{eq:var}
\end{equation}
while the ambient derivatives with respect to the components $\psi_{q,y}$ and $\psi_{q,z}$ give $\tfrac{1}{2} \, 2^{-|N(q)|}$, and $z_q$ does not depend on $\psi_{q,x}$. The variance depends on the degree of $q$ in $E$ and not on $n$. Because $E$ is a ring plus one chord layer, $|N(q)| = 4$ at every width, so \eqref{eq:var} is $1/48$ at 16, 32 and 64 qubits.
\end{theorem}

The proof substitutes \eqref{eq:readout} into each derivative. Each independent angle contributes $\mathbb{E}[\cos^2] = \mathbb{E}[\sin^2] = 1/2$, which gives the factor $2^{-|N(q)|}$, and the uniform measure on $S^2$ contributes $\mathbb{E}[\psi_{q,y}^2] = \mathbb{E}[\psi_{q,z}^2] = 1/3$ with $\mathbb{E}[\psi_{q,y}\psi_{q,z}] = 0$, which gives the factor $1/3$. The angular moments alone would give $1/2$, so the spherical moments are what set the prefactor. Thus, widening the register adds coordinates without reducing the initial gradient scale of an individual coordinate, which supports the width sweep in Fig.~\ref{fig:scaling}. A sufficiently expressive ansatz under the conditions of Ref.~\cite{larocca2025barren} exhibits variance that decays exponentially in $n$, while HyperQ ties the decay to local degree.

The same two-body family loses this property under all-to-all connectivity. There, $|N(q)| = n-1$, so the variance is $\tfrac{1}{3}2^{-(n-1)}$, or $4 \times 10^{-20}$ at 64 qubits. Within this family, increasing degree causes the plateau. We verified \eqref{eq:readout} and \eqref{eq:var} against brute-force statevector calculations at $n = 4$ to $8$.

The result applies to circuit-coordinate gradients at initialization and does not explain every part of model optimization. Because the generators in \eqref{eq:gens} commute, their commutators introduce no new generators, and the dynamical Lie algebra of the interior alone has dimension $n + |E| = 3n$, the readout rotations $U(\psi)$ being excluded. Polynomial Lie-algebra dimension can coincide with both the absence of barren plateaus and efficient classical simulation \cite{ragone2024lie, fontana2024adjoint, cerezo2025simulability}. The constant variance in \eqref{eq:var} and the $\Theta(n)$ readout therefore follow from the same structural restriction. The theorem does not characterize the frozen backbone that consumes the readouts, so it explains why the readout Jacobian avoids width-dependent suppression at fixed degree but does not explain why downstream accuracy increases.

\subsection{Circuit emission hypernetwork}\label{subsec:m-qnas}

HyperQ maps each token representation continuously to coordinates on the shared circuit skeleton with a conditional hypernetwork~\cite{ha2016hypernetworks} implemented as a low-rank adapter~\cite{hu2021lora}. Related hypernetworks emit adapter weights for language models~\cite{phang2023hypertuning, charakorn2025t2l} or circuit parameters for variational algorithms~\cite{qi2025tensorhyper}. Fig.~\ref{fig:overview}, panel (b), illustrates one hypernetwork mapping each token to its own coordinates.

The branch reads one token at a time and carries no attention of its own. Context still
reaches the circuit because $h$ is already contextual when the branch reads it, and the added cost
stays linear in sequence length while the backbone's own attention cost is unchanged. Formally, for $h \in \mathbb{R}^d$ with $d = 2048$, a normalisation and a single hidden
non-linearity produce a rank-$r$ code, and three heads read that code:
\begin{align}
b &= \phi\bigl(W_A\,\mathrm{LN}(h)\bigr) \in \mathbb{R}^{r},
   && \phi = \mathrm{GELU},
   \label{eq:hyper}\\[2pt]
\alpha &= \pi\tanh(W_{\alpha}b) \in \mathbb{R}^{n},
   && \omega_E = \omega_{\max}\tanh(W_{\omega}b) \in \mathbb{R}^{|E|},
   \quad \omega_{\max} = 1.2,
   \label{eq:hyper-angles}\\[2pt]
\psi_q &= \frac{W_{\psi,q}\,b}{\lVert W_{\psi,q}\,b\rVert} \in S^{2},
   && \xi_i = (\alpha,\omega_E,\psi),
   \label{eq:hyper-axes}\\[2pt]
\Delta &= \tfrac{s}{r}\,W_B\,z \in \mathbb{R}^{d_{qkv}},
   && z = (z_1,\dots,z_n),\quad d_{qkv} = 2560.
   \label{eq:hyper-residual}
\end{align}
Here $h = h_i$ is the input of the block's fused query, key and value projection for token $i$%
, $\mathrm{LN}$ a layer normalisation without affine parameters, and $\phi$ the single hidden non-linearity. The maps are bias-free, with $W_A \in \mathbb{R}^{r \times d}$, $W_\alpha \in \mathbb{R}^{n \times r}$, $W_\omega \in \mathbb{R}^{|E| \times r}$, $W_{\psi,q} \in \mathbb{R}^{3 \times r}$ and $W_B \in \mathbb{R}^{d_{qkv} \times n}$, and $s$ is the fixed adapter scale. The axis head emits $3n$ raw scalars that normalise to $n$ points of $S^2$, each carrying two intrinsic coordinates. We initialise $W_B$ at zero, so the branch starts as the identity and the frozen backbone is recovered exactly before training. Here $\xi_i$ holds the complete parameters for token $i$, $z$ is the readout of \eqref{eq:readout}, and $\Delta$ is added to the block's fused query, key and value tensors. Fig.~\ref{fig:hyper} draws this path. With $d = 2048$, $d_{qkv} = 2560$ and $r = 16$, one branch holds $dr + 6nr + d_{qkv}n$ weights, which is $75{,}264$, $117{,}760$ and $202{,}752$ at 16, 32 and 64 qubits. The up-projection $W_B$ carries most of that growth, so a branch is $2.7$ times larger at 64 qubits than at 16. Across $22$ blocks this is $1.7$ to $4.5$ million weights against a $1.1$-billion-parameter backbone. %

\begin{figure}[!tp]
\centering
\noindent\makebox[\textwidth][c]{
\begin{tikzpicture}[
  >={Latex[length=1.5mm]},
  font=\fontsize{7.5}{9}\selectfont,
  node distance=0pt,
  bx/.style={rounded corners=1.8pt, align=center, inner sep=3pt, thin,
             minimum height=0.52cm, minimum width=0.62cm,
             font=\fontsize{7.5}{9}\selectfont},
  sig/.style={bx, draw=QdBackbone, fill=QdSoft},
  hyp/.style={bx, draw=QdAdapter, fill=QdAdapter!14},
  qnt/.style={bx, draw=QdQuantum, fill=QdQuantum!16, minimum width=1.35cm},
  hd/.style={bx, draw=QdAdapter!85, fill=white, minimum height=0.46cm,
             minimum width=0.72cm, font=\fontsize{7}{8}\selectfont},
  fl/.style={->, semithick, black!62, shorten >=0.5mm, shorten <=0.5mm},
  pl/.style={semithick, black!62},
  dim/.style={font=\fontsize{7}{8}\selectfont, text=black!55, inner sep=1.2pt},
  edg/.style={font=\fontsize{7}{8}\selectfont, text=black!70, inner sep=1.2pt},
]

\node[sig]                              (h)   at (0,0)   {$h$};
\node[hyp, right=0.46cm of h]           (ln)             {LN};
\node[hyp, right=0.40cm of ln]          (wa)             {$W_A$};
\node[hyp, right=0.38cm of wa]          (phi)            {$\phi$};
\node[sig, right=0.40cm of phi]         (b)              {$b$};

\node[hd, right=0.52cm of b, yshift=0.80cm]  (ha) {$W_\alpha$};
\node[hd, right=0.52cm of b]                 (hw) {$W_\omega$};
\node[hd, right=0.52cm of b, yshift=-0.80cm] (hp) {$W_\psi$};

\node[qnt, right=1.72cm of hw]          (circ)           {$C(\alpha,\omega)$};
\node[sig, right=0.46cm of circ]        (z)              {$z$};
\node[hyp, right=0.40cm of z]           (wb)             {$W_B$};
\node[sig, right=0.40cm of wb]          (d)              {$\Delta$};

\draw[fl] (h) -- (ln);  \draw[fl] (ln) -- (wa);
\draw[fl] (wa) -- (phi); \draw[fl] (phi) -- (b);

\coordinate (bx) at ($(b.east)+(0.22cm,0)$);
\draw[pl] (b.east) -- (bx);
\foreach \t in {ha,hw,hp}{
  \draw[pl] (bx) |- ($(\t.west)+(-0.18cm,0)$);
  \draw[fl] ($(\t.west)+(-0.18cm,0)$) -- (\t.west);
}

\coordinate (cx) at ($(circ.west)-(0.40cm,0)$);
\foreach \t/\lab/\pos in {%
    ha/{$\alpha\!\in\!\mathbb{R}^{n}$}/above,%
    hw/{$\omega_E\!\in\!\mathbb{R}^{|E|}$}/above,%
    hp/{$\psi\!\in\!(S^{2})^{n}$}/above}{
  \draw[pl] (\t.east) -- node[edg, \pos, pos=0.46, inner sep=1.6pt] {\lab} ($(\t.east)+(1.30cm,0)$);
  \draw[pl] ($(\t.east)+(1.30cm,0)$) -| (cx);
}
\draw[fl] (cx) -- (circ.west);

\draw[fl] (circ) -- (z); \draw[fl] (z) -- (wb); \draw[fl] (wb) -- (d);

\node[dim, below=1pt of h]    {$\mathbb{R}^{d}$};
\node[dim, below=1pt of b]    {$\mathbb{R}^{r}$};
\node[dim, below=1pt of z]    {$\mathbb{R}^{n}$};
\node[dim, below=1pt of d]    {$\mathbb{R}^{d_{qkv}}$};
\node[dim, above=1pt of circ] {$|E| = 2n$};

\begin{scope}[on background layer]
  \node[rounded corners=2.4pt, draw=QdAdapter!55, dashed, thin, inner sep=4pt,
        fit=(ln)(wa)(phi)(b)(ha)(hp)(wb)] {};
\end{scope}
\end{tikzpicture}

}
\caption{Circuit-emission path for one token in HyperQ. A layer normalisation, a
projection to rank $r$ and one hidden non-linearity $\phi$ produce the code $b$
of \eqref{eq:hyper}. Three heads read $b$ and emit the single-qubit angles
$\alpha$, the two-qubit rotation angles $\omega_E$ on the fixed edge set $E$ and the
readout axes $\psi$. The emitted circuit $C(\alpha,\omega)$ returns the $n$
expectation values $z$, which $W_B$ projects into the block's $d_{qkv}$-wide fused query, key and value residual $\Delta$.
Dashed outline, the added branch. Only $h$ and $\Delta$ meet the frozen backbone, and the trainable weights are $W_A$, the three heads and $W_B$.}\label{fig:hyper}
\end{figure}

The single-qubit angles $\alpha$ encode the hidden state over one complete rotation range, preventing drift to multiples of $2\pi$. Each $\omega_e$ controls one of the $|E|$ fixed edges, with zero making it inert, while $\omega_{\max}=1.2$ keeps couplings strictly inside $(-\pi/2,\pi/2)$. Values beyond $\pi/2$ return to an equivalent represented range because \eqref{eq:readout} is periodic in each coupling. Row-normalised $\psi$ adapts the measurement after the Hadamard sandwich while preserving \eqref{eq:iqp}. Each axis has two independent coordinates because $u_q$ is identifiable only up to a right $Z$ rotation and the $x$ component is absent from \eqref{eq:readout}, making a third coordinate unidentifiable. Brute-force statevectors on general two-body edge sets at $n = 2$ to $6$ verified that conjugating the diagonal block with $U(\psi)$ or inserting a free per-qubit axis between the Hadamard layers leaves the IQP family except at isolated axis values. Fig.~\ref{fig:pipeline}c locates each coordinate.

The coordinate space has $n+|E|+2n=5n$ intrinsic dimensions, emitted as $n+|E|+3n$ raw outputs, giving $80$, $160$ and $320$ emitted values at 16, 32 and 64 qubits. %
Since $\omega_e=0$ makes the gate on edge $e$ the identity, a circuit with two-qubit gates on any subset of $E$ is the point of this parameter space where the remaining $\omega_e$ vanish. Generating continuous angles therefore also generates the effective connectivity of each token's circuit, without a discrete selection step.

For Fig.~\ref{fig:motifs}, a gate is active when its generated angle has magnitude at least $0.10$ radians. A generated angle is exactly zero only when the matching row of $W_\alpha$ or $W_\omega$ is orthogonal to $b$, so the summary needs a threshold. For an inactive two-qubit gate, \eqref{eq:readout} multiplies the readouts of both its qubits by $\cos\omega_e \in (0.995, 1]$, and four inactive two-qubit gates at one qubit leave a factor above $\cos^4 0.10 = 0.980$. For a single-qubit gate, write $a_q=\psi_{q,z}\cos\alpha_q-\psi_{q,y}\sin\alpha_q$ for the factor of \eqref{eq:readout} that multiplies the cosines. Changing $\alpha_q$ from zero changes $a_q$ by $\psi_{q,z}(\cos\alpha_q - 1) - \psi_{q,y}\sin\alpha_q$, whose magnitude is at most $2\sin(|\alpha_q|/2) < 2\sin 0.05 < 0.10$ by the Cauchy-Schwarz inequality. The same threshold is therefore a looser criterion for single-qubit gates than for two-qubit gates. A gate group, the ring layer, the chord layer or the single-qubit gates, is dense when at least half of its gates are active. This half-of-group rule, like the threshold value, is a convention for summarising a token's $3n$ angles in one label. Both dense two-qubit layers define Ring and chord, while either alone defines Ring or Chord. With neither two-qubit layer dense, dense single-qubit gates define Product and otherwise Identity. A Product or Identity token can therefore still have active gates on fewer than half of the gates in each group. Each representative token has active ring-edge and chord-edge counts closest to its class medians. The shares use $4{,}096$ tokens at each width.

\subsection{Baseline circuits}\label{subsec:m-baselines}

Each baseline of Tables~\ref{tab:ablation} and~\ref{tab:ensemble} sits in the same quantum branch as HyperQ, keeps the up-projection $W_B$ and follows the same frozen-backbone protocol. It replaces HyperQ's hypernetwork and circuit with its own input map and circuit. Each hand-designed circuit has a fixed layout whose gate angles are trained and shared by all tokens. The token enters only through $n$ input angles $\chi = (\chi_1,\dots,\chi_n)$, which a trained projection computes from $h$. Each circuit starts from $|0\rangle^{\otimes n}$ and returns $\langle Z_q\rangle$ for every qubit $q$. Every circuit except re-upload first applies $R_x(\chi_q)$ to each qubit $q$, and each repeats its layer a fixed number of times. In rx-chain, a layer applies a trained $R_x$ on every qubit followed by CNOT gates between neighbouring qubits. In brickwall, a layer applies two-qubit bricks on alternating even and odd neighbouring pairs, and each brick applies trained $R_y$ gates to both qubits, a CNOT and two further trained $R_y$ gates. In re-upload, a layer applies $R_x(\lambda_q\chi_q + \kappa_q)$ and then $R_y(\nu_q)$ to every qubit, followed by CNOT gates between neighbouring qubits, with $\lambda_q$, $\kappa_q$ and $\nu_q$ trained in each layer. The input angles therefore enter every layer. In iqp-diag, a layer applies Hadamard gates on every qubit, controlled-$R_z$ gates with trained angles between neighbouring qubits, a trained $R_x$ on every qubit and Hadamard gates again. A test-time ensemble combines hand-designed circuits trained separately, each in its own quantum branch. At evaluation, the trained branches are loaded together unchanged, and their outputs are summed into the same query, key and value tensors.

Our token-conditioned differentiable architecture-search baseline~\cite{zoph2017nas, wang2022quantumnas, zhang2022dqas} assigns each token $L=4$ architecture-search slots, where $L$ differs from the sequence length in Section~\ref{subsec:m-diffusion}. The slots lie between the two Hadamard layers, and the motif chosen for a slot is the gate group that the slot runs. The 3-motif variant selects among single-qubit $Z_q$, ring $Z_iZ_j$ and chord-layer $Z_iZ_j$ generators, while the 6-motif variant adds three classes outside the IQP family, including a $Y$ rotation. Every gate of every motif in every slot has its own trained angle, shared by all tokens, and the token also enters through $n$ input angles computed from $h$ by a trained projection, as in the hand-designed circuits. For slot $s$ with a catalogue of $M$ classes, $\ell_s=W_sb\in\mathbb{R}^{M}$ and $m_s=\arg\max_c\ell_{s,c}$, while the straight-through backward pass uses $\mathrm{softmax}(\ell_s/\tau)$ with $\tau=1$. This softmax is the motif probability $p$ of Fig.~\ref{fig:search}a, and it starts near $1/M$ because $W_s$ starts from small random weights. This gives $3^4=81$ and $6^4=1296$ architectures per token on $E$, trained only through \eqref{eq:mdm} without an auxiliary search reward or separate validation loop, as reported in Fig.~\ref{fig:search}a.

\subsection{Circuit synthesis and training}\label{subsec:m-training}

All tokens use the strict-IQP topology from Eq.~\eqref{eq:iqp}, comprising $\alpha$ single-qubit $Z$ rotations and $|E|$ two-body $Z_iZ_j$ rotations at $\omega_E$ between two $H^{\otimes n}$ layers. Couplings compile to controlled-$Z$ rotations on $E$, while commutativity in Eq.~\eqref{eq:gens} permits parallel packing. Routing $E$ once per width fixes layout, two-qubit gate count and depth across tokens. At optimisation level two, backend-target transpilation updates the $Z$ observables for the layout and batches all circuits in one Qiskit Runtime EstimatorV2 job. Each hardware run requires every $\langle Z_q\rangle$ to agree with the tensor-network contraction within $10^{-4}$.

Through the supervised \texttt{sft\_lora\_mdm} pipeline, we freeze LLaDA-1.1B and optimize only the low-rank quantum branch and circuit heads attached to all $22$ transformer blocks. Under grouped-query attention, the adjacent fused query-key-value projection width $2048+2\times256=2560$ sets the branch adapter output dimension. Runs use HuggingFace \texttt{accelerate} with bf16 mixed precision on one node of eight NVLink-connected NVIDIA A100 $80$\,GB SXM accelerators, AdamW with $\beta_1=0.9$, $\beta_2=0.95$ and zero weight decay, a cosine schedule with 24-step warm-up to $2\times10^{-4}$, clipping at $1.0$, $8$ sequences per device accumulated over $4$ batches, three epochs from seed $3407$, LoRA rank $r=16$, scaling $s=32$, dropout $0.05$, and checkpoints every $100$ steps, giving effective batch size $256$. HyperQ's $20{,}000$-pair subset yields $78$ optimizer steps per epoch and $235$ total. At $L=1024$, each run processes $6.1\times10^7$ tokens, one tenth of the baselines' $6.1\times10^8$. The sole signal is the masked cross-entropy diffusion bound in \eqref{eq:mdm} at masked positions.

Activation gradients through the frozen backbone's $N$ parameters cost approximately $4N$ floating-point operations per token instead of $6N$ with weight gradients, while the branch retains forward, activation-gradient and weight-gradient passes. We estimate backbone compute at $30$ per cent of each device's $312$\,TFLOP/s bf16 peak, including dataloading, checkpointing and optimizer overhead, and charge the bandwidth-bound branch against $2.04$\,TB/s memory bandwidth.

The exact closed form in \eqref{eq:readout} costs $\Theta(n)$ forward and backward because $|E|=2n$, with prefix and suffix products avoiding division by near-zero cosines. For $a_q=\psi_{q,z}\cos\alpha_q-\psi_{q,y}\sin\alpha_q$,
\[
\partial z_q/\partial\omega_{qj}
=
-\,a_q\sin\omega_{qj}
\prod_{k\in N(q),\,k\neq j}\cos\omega_{qk}.
\]
A coupling near $\pi/2$ suppresses the other three gradients. At $\omega=\pi/2$, their mean magnitude falls from $0.385$ to $3.7\times10^{-5}$ with those couplings at $0.6$ radians and single-qubit amplitude one. Squashing at $\omega_{\max}=1.2$ leaves fewer than $0.2$ per cent suppressed, versus $3.2$ per cent when the range reaches $\pi/2$.

The fp32 prefix and suffix tables contain $|E|$ entries and require $8|E|+16n$ bytes per token per block with other intermediates. Scaling from $16$ to $32$ or $64$ qubits multiplies this by $2.00$ or $4.00$, versus $4.13$ and $16.8$ for a complete graph. Bounded degree keeps mean readout magnitude at $0.16$ on $E$ at every width, whereas $n-1$ cosine factors yield $2.6\times10^{-13}$ at $64$ qubits for couplings uniform on $(-\pi,\pi)$. All models share the LoRA rank, learning-rate schedule and frozen-backbone protocol, so cost varies only with register width and branch type. We measure latency and activation memory on the eight-A100 node.

TensorCircuit-NG with its JAX backend evaluates the closed form on the training GPUs. At $8$ to $16$ qubits, it must agree within $10^{-4}$ with a batched \texttt{complex64} statevector simulator and exact \texttt{float64} tensor-network contraction. Agreement at 16 qubits validates the exact bounded-treewidth tensor network at $32$ and $64$ qubits. Fig.~\ref{fig:hw} uses the analytic reference and tensor-network offline control. A pure array implementation is traced once per width, XLA dispatches Python only on the first call, \texttt{vmap} batches circuits, and fixed-size padded chunks bound memory while keeping traced shapes constant.

\subsection{Datasets and evaluation protocols}\label{subsec:m-eval}

A comparison between HyperQ and classical adaptation must control training data, preprocessing and downstream scoring so that the circuit branch is the principal experimental difference. We follow the supervised fine-tuning setting reported for the LLaDA 8B model~\cite{nie2025llada}. The classical baselines are fine-tuned on $200{,}000$ prompt-and-response pairs. HyperQ uses a uniformly sampled $20{,}000$-pair subset containing one tenth of the baseline training tokens.

The inherited preprocessing leaves prompt tokens uncorrupted and masks response tokens independently. It applies \eqref{eq:mdm} only at masked response positions. Padding end-of-sequence tokens remain part of the response, which trains the model to stop without reconstructing the instruction. Multi-turn dialogues are split into single-turn pairs, and training lasts three epochs with a global batch of $256$ pairs.

We match the inherited objective, padding convention, three-epoch duration and global batch. Three settings change to support frozen-backbone circuit adaptation. \textit{(i)} LLaDA updates all parameters with weight decay $0.1$. Its learning rate warms linearly to $2.5\times10^{-5}$ over the first $50$ iterations, remains constant for an intermediate interval and decays linearly to $2.5\times10^{-6}$ during the final tenth of training. We update only the rank-$16$ branch and its two circuit heads using Section~\ref{subsec:m-training}. \textit{(ii)} LLaDA sets sequence length dynamically within each minibatch and reports no maximum. We fix the block cap at $L = 1024$ tokens so every step contains the same number of per-token circuits. \textit{(iii)} LLaDA reports supervised fine-tuning only for its 8B model. The pair counts here are therefore our chosen experimental scales and do not reproduce a published LLaDA run.

We evaluate zero-shot downstream quality with the \texttt{eval\_mdm} pipeline, an lm-eval-style harness adapted to masked-diffusion readout. For each multiple-choice item, it scores every candidate completion by conditional log-likelihood under the diffusion bound. The estimate uses up to 128 Monte Carlo mask samples per item and an evaluation batch size of 8.

The complete suite contains HellaSwag, ARC-e, ARC-c, WinoGrande, PIQA, OpenBookQA, MMLU, BoolQ, RACE, GSM8K, LAMBADA (cloze) and TruthfulQA (mc2). The summary value averages six benchmarks: ARC-e, HellaSwag, PIQA, BoolQ, RACE and GSM8K. Table~\ref{tab:main} compares HyperQ with two classical baselines of equal parameter budget, the masked-diffusion LLaDA-1.1B and the autoregressive TinyLlama-1.1B.

Hardware evaluation measures how much downstream quality changes when the analytic circuit readout is replaced by a noisy-device readout. We execute emitted per-token circuits on a $156$-qubit IBM superconducting processor with Qiskit Runtime EstimatorV2 at $10\,000$ shots. This count targets a precision of $10^{-2}$ for every expectation value consumed by the backbone.

\backmatter

\bibliography{claudetodo,references,custom}

\end{document}